\documentclass[a4paper,onecolumn, 10pt]{article} 

\usepackage{amsmath,amssymb}
\usepackage[utf8]{inputenc}
\usepackage{authblk}
\usepackage{filecontents}

\usepackage{abstract}
\usepackage[top=1in, bottom=1in, left=1in, right=1in]{geometry}
\usepackage{hyperref} 
\usepackage[backend=biber, style=nature, sorting=none]{biblatex}
\usepackage{color}
\usepackage{xspace}
\newcommand{\red}[1]{\textcolor{black} {#1}\xspace}

\usepackage{textcomp}
\usepackage{setspace}
\usepackage{tabularx}
\usepackage{graphicx}
\usepackage{layout}
\usepackage[margin=10pt,font=small,labelfont=bf,format=plain]{caption}
\renewcommand{\figurename}{Fig.}
\let\svthefootnote\thefootnote
\newcommand\freefootnote[1]{%
  \let\thefootnote\relax%
  \footnotetext{#1}%
  \let\thefootnote\svthefootnote%
}

\newcommand{\beginsupplement}{%
        \setcounter{table}{0}
        \renewcommand{\thetable}{\arabic{table}}%
        \setcounter{figure}{0}
        \renewcommand{\thefigure}{\arabic{figure}}%
        \renewcommand{\figurename}{Supplementary Fig.}
        \renewcommand{\tablename}{Supplementary Table}
     }

\title{Nonequilibrium Charge-Density-Wave Order Beyond the Thermal Limit} 

\author[1]{J. Maklar}
\author[1]{Y. W. Windsor}
\author[1,$^*$]{C. W. Nicholson}
\author[1,$^\dagger$]{M. Puppin}
\author[2,3]{P. Walmsley}
\author[3,4]{V. Esposito}
\author[4]{M. Porer}
\author[4]{J. Rittmann}
\author[5]{D. Leuenberger}
\author[6]{M. Kubli}
\author[6]{M. Savoini}
\author[6]{E. Abreu}
\author[6]{S. L. Johnson}
\author[4]{P. Beaud}
\author[4]{G. Ingold}
\author[4]{U. Staub}
\author[2,3]{I. R. Fisher}
\author[1]{R. Ernstorfer}
\author[1]{M. Wolf}
\author[1]{L. Rettig}

\affil[1]{Fritz-Haber-Institut der Max-Planck-Gesellschaft, Faradayweg 4-6, D-14195 Berlin, Germany}
\affil[2]{Geballe Laboratory for Advanced Materials and Department of Applied Physics, Stanford University, CA 94305, USA}
\affil[3]{Stanford Institute for Materials and Energy Sciences, SLAC National Accelerator Laboratory, 2575 Sand Hill Road, Menlo Park, CA 94025, USA}
\affil[4]{Swiss Light Source, Paul Scherrer Institut, CH-5232 Villigen PSI, Switzerland}
\affil[5]{Department of Physics, University of Zürich, CH-8057 Zürich, Switzerland}
\affil[6]{Institute for Quantum Electronics, Physics Department, ETH Zürich, CH-8093 Zürich, Switzerland}

\date{\today}

\begin{document}
\maketitle 
    \begin{abstract}
The interaction of many-body systems with intense light pulses may lead to novel emergent phenomena far from equilibrium. Recent discoveries, such as the optical enhancement of the critical temperature in certain superconductors and the photo-stabilization of hidden phases, have turned this field into an important research frontier. Here, we demonstrate nonthermal charge-density-wave (CDW) order at electronic temperatures far greater than the thermodynamic transition temperature. Using time- and angle-resolved photoemission spectroscopy and time-resolved X-ray diffraction, we investigate the electronic and structural order parameters of an ultrafast photoinduced CDW-to-metal transition. Tracking the dynamical CDW recovery as a function of electronic temperature reveals a behaviour markedly different from equilibrium, which we attribute to the suppression of lattice fluctuations in the transient nonthermal phonon distribution. A complete description of the system's coherent and incoherent order-parameter dynamics is given by a time-dependent Ginzburg-Landau framework, providing access to the transient potential energy surfaces.
    \end{abstract}

\freefootnote{Current address: \\$^*$Department of Physics and Fribourg Center for Nanomaterials, University of Fribourg, Chemin du Musée 3, CH-1700 Fribourg, Switzerland\\
$^\dagger$Laboratory of Ultrafast Spectroscopy, ISIC, Ecole Polytechnique Fédérale de Lausanne (EPFL), CH-1015 Lausanne, Switzerland}

\section*{Introduction}
Complex solids exhibit a multitude of competing and intertwined broken symmetry states originating from a delicate interplay of different degrees of freedom and dimensionality. Among these states, charge-density-waves (CDWs) are a ubiquitous phase characterised by a cooperative periodic modulation of the charge density and of the crystal lattice, mediated by electron-phonon coupling\supercite{motizuki1986, gruner1994, pouget2016}. While lattice and charges are intrinsically coupled in equilibrium, ultrafast optical excitation allows to selectively perturb each of these subsystems and to probe the melting of order and its recovery as a real-time process. This approach grants access to the relevant interactions of CDW formation\supercite{demsar1999, perfetti2006, schmitt2008, eichberger2010, mohr2011, hellmann2012time, sohrt2014fast, huber2014, porer2014, rettig2016, monney2016revealing, yang2020bypassing}, to out-of-equilibrium and metastable states\supercite{tsuji2013, stojchevska2014, zhang2016, gerasimenko2019} and elucidates competing orders\supercite{fausti2011, wandel2020, kogar2020}. 

In close analogy to superconductivity, the formation of a CDW broken symmetry ground state can be described by an effective mean field that serves as an order parameter, which is governed in equilibrium by a static free energy surface. While mean field theory captures the phase transition on a qualitative level, thermal lattice fluctuations reduce the critical temperature $T_\mathrm{c}$ of long-range 3D order significantly below the predicted mean field value $T_{\mathrm{MF}}$\supercite{motizuki1986, gruner1994}. It is of strong interest how our understanding of phase transitions in the adiabatic limit can be adapted to a non-equilibrium, dynamical setting induced by an impulsive excitation\supercite{huber2014, beaud2014, wall2018ultrafast, nicholson2018beyond, neugebauer2019, dolgirev2020}. It remains an open question whether the thermal transition temperature is still a relevant quantity in the description of such an out-of-equilibrium state, and which parameters permit transient control of $T_{\mathrm{c}}$\supercite{fausti2011, mitrano2016possible, singer2016photoinduced, nicholson2016ultrafast, cavalleri2018photo, tengdin2018critical}.

Symmetry-broken phases also allow for collective excitations of the order parameter, as observed in a variety of systems, including CDW compounds, superconductors and atoms in optical lattices\supercite{endres2012, matsunaga2013, torchinsky2013}. Two types of modes emerge in the symmetry-broken ground state, related to a variation of the amplitude and the phase of the complex order parameter, i.e., the Higgs amplitude mode (AM) and the Nambu-Goldstone phase mode. In CDW compounds, upon impulsive excitation, the AM manifests as coherent oscillations of the electronic and structural order-parameter amplitudes\supercite{demsar1999, yusupov2008, schmitt2008}. However, recent studies investigating the structural dynamics of various CDW compounds upon strong perturbation hint towards collective modes at increased frequencies far beyond the intrinsic AM\supercite{huber2014, neugebauer2019, trigo2019}.

To address these issues, we investigate the electronic and structural order of optically excited bulk TbTe$_3$, a prototypical CDW compound of the rare-earth tritelluride family\supercite{ru2006, brouet2008}. Using time- and angle-resolved photoemission spectroscopy (trARPES) in combination with time-resolved X-ray diffraction (trXRD), schematically depicted in Fig.\,\ref{fig:setup}a, we extract the amplitude of the electronic and structural order parameters and the electronic temperature as functions of pump-probe delay $t$. This reveals CDW formation at electronic temperatures substantially above the thermal critical temperature. We attribute this transient stabilization to a reduced contribution of lattice fluctuations in the out-of-equilibrium state due to a nonthermal phonon population. Furthermore, with increasing excitation density, the coherent order parameter dynamics indicate a transition from the AM regime to a high-frequency regime, driven by a modification of the underlying potential energy surface. We model the order-parameter dynamics in a time-dependent Ginzburg-Landau framework, which further supports the scenario of a nonthermal stabilization of the CDW order.

\section*{Results}
\subsection*{Electronic and structural CDW signatures}
\begin{figure}[!ht]
\centering
\includegraphics[width=0.5\columnwidth]{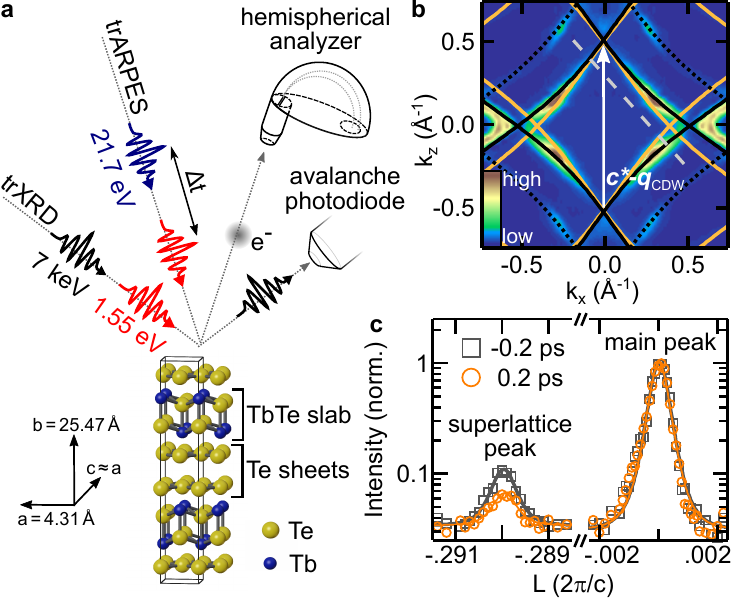}
\caption{\textbf{Experimental scheme.} (\textbf{a}) Schematic of the trARPES and grazing-incidence trXRD experiments. TbTe$_3$ is a quasi-2D compound consisting of a stack of Te sheets and TbTe slabs. (\textbf{b}) Symmetrized FS of TbTe$_3$ ($T=100$\,K, $t=0$\,fs). Below $T_{\mathrm{c}}$, the spectral weight within the nested FS regions connected by the CDW wave vector $\boldsymbol{c}^*\mbox{-}\boldsymbol{q}_\mathrm{CDW}$ vanishes\supercite{brouet2008}. The black solid and dotted lines correspond to Te 5p\textsubscript{x}/5p\textsubscript{z} bands from tight-binding calculations. FS nesting also leads to the formation of shadow bands (orange lines). The grey dashed line indicates the momentum-direction analysed in Figs.\,\ref{fig:shortdelays}a-c. (\textbf{c}) Representative X-ray Bragg peaks with Voigt fits along the (3 7 L) direction before and after optical excitation (absorbed fluence $F=1.35$\,mJ cm$^{-2}$).}
\label{fig:setup}
\end{figure}

First, using ARPES, we analyse the Fermi surface (FS) of TbTe$_3$ at $T=100$\,K, well below $T_{\mathrm{c}}=336$\,K, the transition temperature of the unidirectional CDW phase\supercite{ru2008}. The electronic properties near $E_{\mathrm{F}}$ are governed by the Te sheets (Fig.\,\ref{fig:setup}a), which give rise to the diamond-shaped bands shown in Fig.\,\ref{fig:setup}b. Strongly wave-vector dependent electron-phonon coupling\supercite{maschek2015}, in conjunction with a moderately well-nested Fermi surface\supercite{laverock2005fermi}, lead to a unidirectional CDW in which some portions of the Fermi surface are gapped while others remain metallic\supercite{brouet2008}. To study the effect of the CDW on the lattice, we investigate the intensity of superlattice (SL) Bragg peaks using trXRD. These SL peaks arise from the periodic lattice distortion associated with the CDW, and are displaced by the CDW wave vector $\pm \boldsymbol{q}_{\mathrm{CDW}}$ from the main peak positions\supercite{overhauser1971observability,ru2008}. As Fig.\,\ref{fig:setup}c shows, photoexcitation strongly suppresses the SL peak corresponding to a rearrangement of the atomic mean positions towards the trivial metallic phase, while the main lattice peak reflecting the average crystal structure shows only minor changes.

\begin{figure*}[!ht]
\centering
\includegraphics[width=1\textwidth]{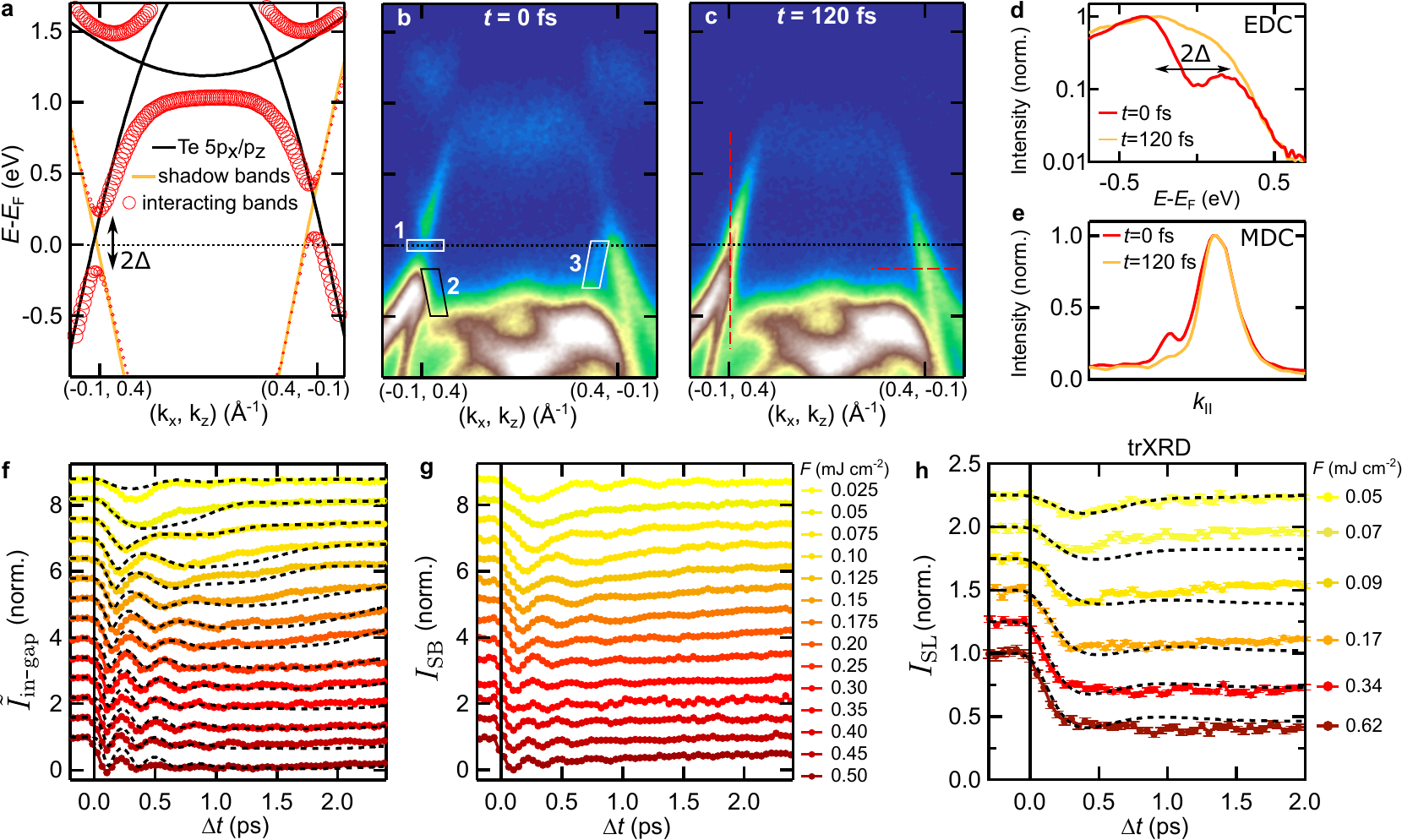}
\caption{\textbf{CDW band structure dynamics.} (\textbf{a}) Tight-binding bands along the momentum-direction indicated by the dashed grey line in Fig.\,\ref{fig:setup}b. The black and orange curves correspond to the non-interacting Te main and shadow bands, respectively. The red circles mark the hybridized bands with interaction potential $\Delta$. The circle size illustrates the spectral weight. (\textbf{b-c}) trARPES measurements ($F=0.45$\,mJ cm$^{-2}$) along the momentum direction shown in \textbf{a}. At $t=0$\,fs, the energy gap at $E_{\mathrm{F}}$ (box 1) and shadow bands (boxes 2, 3) indicate the CDW order. After 120\,fs, the CDW vanishes, and the energy gap and shadow band intensity are strongly suppressed. (\textbf{d-e}) Energy and momentum distribution curves along the dashed vertical and horizontal lines in \textbf{c}, respectively. (\textbf{f}) Inverted in-gap intensity $\tilde{I}_{\mathrm{in\mbox{-}gap}}=1-I_{\mathrm{in\mbox{-}gap}}$ with in-gap intensity $I_{\mathrm{in\mbox{-}gap}}$ (box 1 in \textbf{b}, normalized by the respective pre-excitation values) as function of pump-probe delay for various fluences (displaced vertically). Normalized time-dependent Ginzburg-Landau simulations are shown in black. For details of the model, see main text and Supplementary Note\,\ref{supp:model}. (\textbf{g}) Normalized shadow band intensity extracted from box 2. The shadow band intensity obtained from box 3 is shown in Supplementary Fig.\,\ref{fig:SB_data}. (\textbf{h}) Time evolution of the (2 10 1+$q_{\mathrm{CDW}}$) SL peak intensity for various fluences (displaced vertically) with layered Ginzburg-Landau simulations, see Supplementary Note\,\ref{supp:tdgl_structural}. The curves are normalized by their respective pre-excitation values. The error bars correspond to one standard deviation from photon counting statistics.}
\label{fig:shortdelays}
\end{figure*}

Next, we investigate the electron dynamics associated with the CDW upon photoexcitation. We focus on an energy-momentum cut that contains the electronic signatures of the CDW, namely the energy gap at $E_{\mathrm{F}}$ in the nested regions and the backfolded shadow bands\supercite{voit2000}, shown in Figs.\,\ref{fig:shortdelays}a-b. At temporal pump-probe overlap ($t=0$\,fs), the interacting tight-binding model introduced by Brouet et al.\supercite{brouet2008} is in excellent agreement with the observed quasiparticle dispersion: In the nested region (left side of Figs.\,\ref{fig:shortdelays}a-b), we observe a pronounced hybridization energy gap at $E_{\mathrm{F}}$. In the imperfectly nested region (right side), the Te band exhibits metallic behaviour, as the energy gap is located above $E_{\mathrm{F}}$. Furthermore, we observe faint shadow bands in the vicinity of the energy gaps (boxes 2 and 3 in Fig.\,\ref{fig:shortdelays}b). Within 120\,fs, the system undergoes a photoinduced CDW-to-metal transition\supercite{schmitt2008}, as apparent from the transient suppression of the energy gap and the shadow bands, see Figs.\,\ref{fig:shortdelays}c-e. 

\subsection*{CDW order-parameter dynamics}
\begin{figure}[!ht]
\centering
\includegraphics[width=0.5\columnwidth]{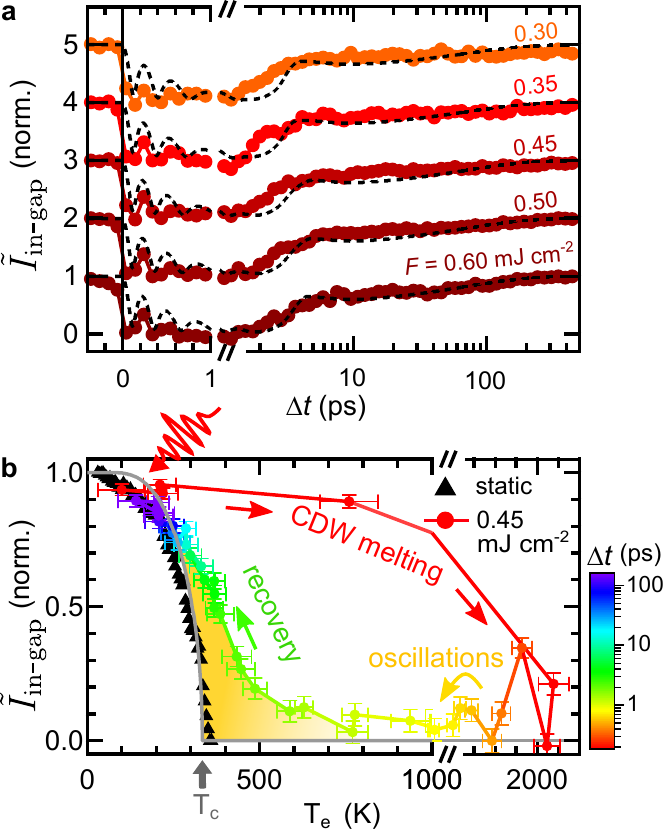}
\caption{\textbf{CDW recovery dynamics.} (\textbf{a}) Time evolution of the inverted in-gap intensity in the high-fluence regime (displaced vertically). Normalized time-dependent Ginzburg-Landau simulations are shown in black. (\textbf{b}) Inverted in-gap intensity versus extracted electronic temperatures. One standard deviation of the $T_\mathrm{e}$ fit (horizontal error bars) and one standard deviation derived from electron counting statistics (vertical error bars) are given as uncertainty. $\tilde{I}_{\mathrm{in\mbox{-}gap}}$ extracted from a static temperature series (black markers, $T$-values from heater setpoints, curve normalized to the lowest accessible $T$-value) is in general agreement with the BCS-type $T$-dependence of the order parameter (grey curve). The dynamic trace shows the full cycle of laser-heating and CDW melting, coherent oscillations and CDW recovery (delay encoded in the color code). The yellow shaded area marks the region of dynamical CDW formation at electronic temperatures above $T_{\mathrm{c}}$. The pre-excitation value of the dynamic trace ($T=100$\,K) is normalized to the corresponding value of the static $T$-dependence.}
\label{fig:longtermrecovery}
\end{figure}

The CDW-to-metal transition can be described by an order parameter $\psi$, with $|\psi|=0$ in the metallic and $0<|\psi| \leq 1$ in the CDW phase. Due to the coupling between charges and lattice, the transition can be characterized by an electronic ($\psi_\mathrm{e}$) or a structural ($\psi_\mathrm{s}$) order parameter. We utilize trARPES to access the amplitude of the electronic order parameter $|\psi_\mathrm{e}|$. Most directly, $|\psi_\mathrm{e}|$ can be extracted by tracking the energy gap $2\Delta$ at $E_{\mathrm{F}}$\supercite{rettig2014,rettig2016}. However, this method faces practical limitations due to the vanishing occupation of bands above $E_{\mathrm{F}}$ after a few 100\,fs and due to the limited experimental energy resolution. Therefore, we choose two alternative metrics to quantify the CDW order: We introduce the inverted in-gap intensity $\tilde{I}_{\mathrm{in\mbox{-}gap}}=1-I_{\mathrm{in\mbox{-}gap}}$ with normalized in-gap intensity $I_{\mathrm{in\mbox{-}gap}}$, extracted from box 1 in Fig.\,\ref{fig:shortdelays}b. We find that this metric -- for the chosen region of interest and our experimental resolution -- follows a BCS-like temperature dependence in equilibrium, as confirmed by static measurements (black markers in Fig.\,\ref{fig:longtermrecovery}b), and thus is considered equivalent to $|\psi_\mathrm{e}|$. Further, as the inverted in-gap intensity is derived from a region where the gap is centered around $E_{\mathrm{F}}$, it is unaffected by thermal changes to the distribution function. As a second metric, we extract the shadow band intensity $I_{\mathrm{SB}} \propto |\psi_\mathrm{e}|$\supercite{voit2000, nicholson2016ultrafast} from box 2 in Fig.\,\ref{fig:shortdelays}b.

Using these equivalent metrics, we investigate the photoinduced CDW suppression and recovery over a wide range of fluences, as shown in Figs.\,\ref{fig:shortdelays}f-g. For a low absorbed fluence of 0.025\,mJ cm$^{-2}$ below the CDW melting threshold, we observe a weak modulation of the CDW gap and SB intensity corresponding to the AM of the CDW at $\omega_{\mathrm{AM}}/2\pi = 2.2$\,THz (see Supplementary Fig.\,\ref{fig:supp_CDW_AM}). At the CDW melting threshold $\approx 0.05$\,mJ cm$^{-2}$, the AM softens and becomes overdamped, while the CDW melting time $t_{\mathrm{melt}}$ slows down, and the energy gap and SB intensity vanish almost completely. Upon crossing the melting threshold, we observe a fast initial quench of the CDW within $t_{\mathrm{melt}} \approx 100$\,fs (see Supplementary Fig.\,\ref{fig:slowdown_melting}), followed by few damped coherent oscillations that exhibit a pronounced frequency reduction with pump-probe delay (down-chirp). Interestingly, the initial frequency of the collective excitation increases with fluence, doubling at the highest accessible fluences. Concurrently, the time required to restore the ground state after perturbation steadily increases with fluence, leading to a persistent suppression of the CDW for a few ps at the highest excitation densities we used.

To gain a complementary view of the photoinduced phase transition, we use trXRD to extract the structural order parameter from the normalized SL peak intensity upon optical excitation\supercite{beaud2014, huber2014, trigo2019}, which, in first approximation, is given by $I_{\mathrm{SL}}(t) \propto |\psi_\mathrm{s}(t)|^2$. As Fig.\,\ref{fig:shortdelays}h shows, the SL response qualitatively resembles the dynamical quench and recovery of the extracted electronic order parameter. In the low-fluence regime, a weak initial suppression is followed by a quick recovery of the SL structure, on top of which a faint modulation can be identified (see Supplementary Fig.\,\ref{fig:supp_X_ray_AM}). In the high-fluence regime, the SL peak intensity is strongly quenched, and, with increasing fluence, the time required to recover diverges. In contrast to the electronic response, we do not observe clear coherent oscillations of the SL peak intensity upon strong excitation. This originates most likely from the lower temporal resolution of the trXRD setup and the contribution of sub-surface crystal layers with varying, lower excitation densities (see Supplementary Note\,\ref{supp:tdgl_structural}). Recent trXRD experiments with improved temporal resolution have revealed fluence-dependent collective excitations of the SL peak intensity in a closely related tritelluride\supercite{trigo2019} -- in agreement with our observations for $\psi_\mathrm{e}$. Furthermore, while the SL intensity $I_{\mathrm{SL}}$ drops linearly with excitation density shortly after excitation, this behaviour plateaus after crossing a fluence of $\approx 0.1$\,mJ cm$^{-2}$. This results in a residual SL intensity of $35\%$ even after strong excitation of up to 1.35\,mJ cm$^{-2}$. We assign this persisting SL background to a contribution of unexcited sample volumes due to surface steps caused by crystal cleaving\supercite{huber2014}. Nonetheless, the trXRD data clearly shows that not only the electronic, but also the lattice superstructure is melted upon strong photoexcitation. The qualitative agreement of the electronic and structural response demonstrates a strong coupling between electronic and lattice degrees of freedom on ultrafast timescales, and suggests an equivalent treatment of $|\psi_\mathrm{s}|$ and $|\psi_\mathrm{e}|$ within the experimental time resolution.

Diffraction also probes the long-range coherence of the SL phase. While phase coherence plays a secondary role in the low-fluence regime, it becomes increasingly important during the CDW recovery after strong perturbation due to the creation of topological defects. These dislocation-type defects broaden the SL peaks, locally decrease the amplitude of the periodic lattice modulation, and can persist long after the CDW amplitude has recovered\supercite{vogelgesang2018phase, zong2019evidence, trigo2020formation}. Therefore, rather than trXRD, we employ trARPES to access the amplitude of the order parameter throughout the full recovery to equilibrium. As shown in Fig.\,\ref{fig:longtermrecovery}a, in the high-fluence regime, the majority of the CDW order is restored after $\approx 5$\,ps, followed by a complete recovery on a 100\,ps timescale.

\subsection*{Transient electronic temperature}
Time-resolved ARPES allows to extract the transient electronic temperatures from Fermi-Dirac fits to the energy distribution of metallic regions of the FS (see Supplementary Note\,\ref{supp:Te}), and thereby to compare the non-equilibrium CDW melting and recovery to the mean field behaviour upon thermal heating. Remarkably, in the dynamic case, the electronic order parameter does not follow the mean field dependence governed by $T_{\mathrm{c}}$. In the low-fluence regime below the CDW melting threshold, electronic temperatures reach up to 500\,K, far above $T_{\mathrm{c}}=336$\,K (see Supplementary Fig.\,\ref{fig:Te_multi}a). Yet, photoexcitation causes only a minor initial suppression of the energy gap and of the periodic lattice distortion, and initiates a collective AM oscillation --  a hallmark of the CDW state. 

In the high-fluence regime, the CDW is fully suppressed ($\tilde{I}_{\mathrm{in\mbox{-}gap}}=I_{\mathrm{SB}}=0$) as initial electronic temperatures exceed 2000\,K. However, recovery of the CDW order already sets in when the electronic system is still at elevated temperatures $T_{\mathrm{e}} \gg T_{\mathrm{c}}$. To illustrate this dynamic behaviour, Fig.\,\ref{fig:longtermrecovery}b presents the inverted in-gap intensity of the melting and the recovery cycle as a function of extracted electronic temperatures. In the out-of-equilibrium setting, CDW order reappears below $T_\mathrm{e}\approx 600$\,K (yellow shaded area), indicating an increased effective critical temperature $T_{\mathrm{c}}^{*}$. At delay times of several ps, corresponding to electronic temperatures of $T_{\mathrm{e}} \leq T_{\mathrm{c}}$, the dynamic behaviour converges to the equilibrium $T$-dependence. This trend of nonthermal CDW recovery is consistent over a wide range of fluences (see Supplementary Fig.\,\ref{fig:supp_ingap_Te}).

\subsection*{Time-dependent Ginzburg-Landau theory}
\begin{figure*}[!ht]
\centering
\includegraphics[width=1\textwidth]{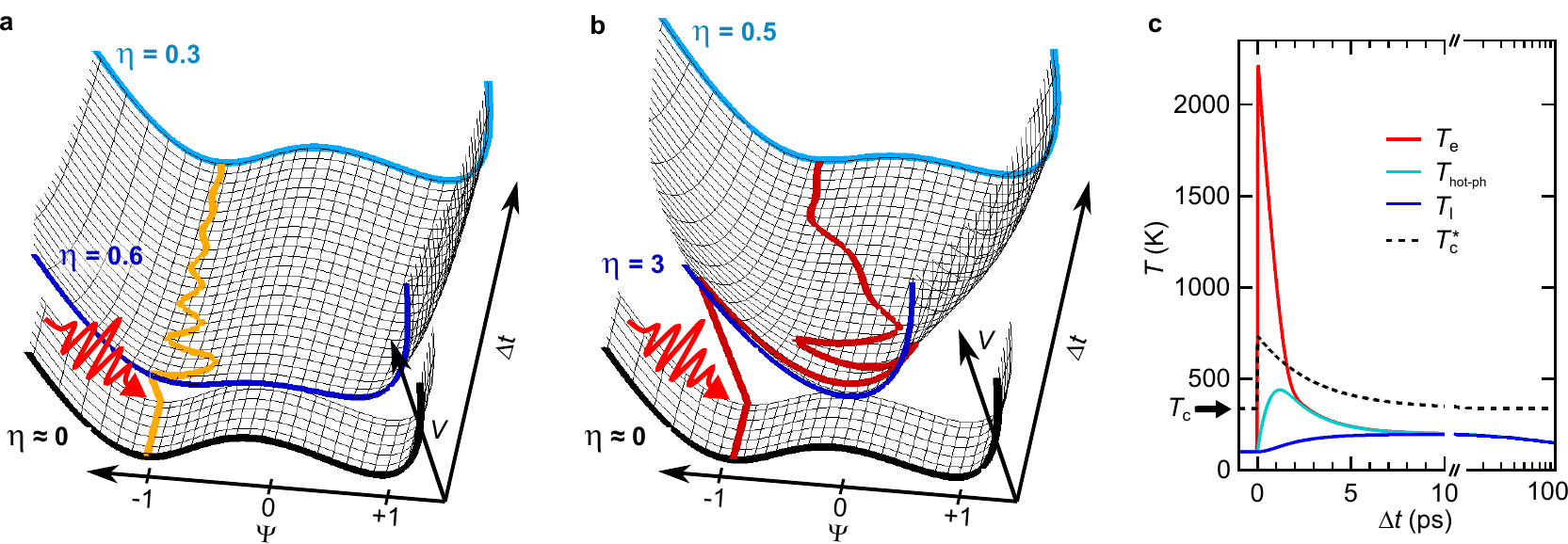}
\caption{\textbf{Simulated order-parameter dynamics and 3TM.} Transient potential energy surface and order-parameter pathway upon (\textbf{a}) weak and (\textbf{b}) strong optical excitation. The potential shapes before excitation (black curve), at 0\,ps (dark blue) and 3.5\,ps (light blue) are highlighted. (\textbf{a}) In the AM regime, the double-well potential is weakly modified, while in (\textbf{b}) the overshoot regime, the CDW melting threshold is reached, resulting in a single-well shaped potential, followed by a relaxation to the double-well ground state. (\textbf{c}) 3TM of electronic, hot phonon and lattice temperatures $T_{\mathrm{e}}$, $T_{\mathrm{hot\mbox{-}ph}}$ and $T_{\mathrm{l}}$ in the regime of strong perturbation ($F$=0.35\,mJ cm$^{-2}$). In the 3TM, the optical excitation of the electronic system is followed by an energy transfer to certain strongly-coupled optical phonons, widely observed in materials with selective electron-phonon coupling\supercite{perfetti2007, tao2013, johnson2017, nicholson2019PRB, storeck2019hot, dolgirev2020}. Subsequently, this hot phonon subset equilibrates with the remaining lattice phonon bath on a ps timescale ($\tau_{\mathrm{ph\mbox{-}ph}}$). To account for the recovery of the base temperature via heat diffusion on a 100\,ps timescale, the lattice is coupled to an external heat sink. The black dashed line indicates the rescaled critical temperature $T_{\mathrm{c}}^{*}$. In the 3TM simulations, material properties of the related compound LaTe$_3$\supercite{dolgirev2020} were used.}
\label{fig:potentials}
\end{figure*}

Near the transition temperature, the order parameter can be approximated by the Landau theory of second-order phase transitions\supercite{gruner1994}. Thus, to simulate the dynamics of the order parameter in TbTe$_3$, we make the following ansatz for the effective potential energy surface (in dimensionless units) based on time-dependent Ginzburg-Landau (tdGL) theory\supercite{yusupov2010, schaefer2014, huber2014, trigo2019, dolgirev2020}:
\begin{equation}
   V(\psi,t)=-\frac{1}{2}\big( 1-\eta(t)  \big) \psi^2 +  \frac{1}{4} \psi^4\,.
   \label{equ:potential}
\end{equation}
Upon perturbation, the dynamics of the order parameter are determined by the equation of motion derived from Eq.\,\ref{equ:potential} (see Supplementary Note\,\ref{supp:model}). The transient modification of the potential, resulting from the laser excitation and subsequent relaxation, is modelled by the ratio of the electronic temperature and the critical temperature $\eta(t)=T_{\mathrm{e}}/ T_{\mathrm{c}}$. Motivated by the increased transient ordering temperature discussed above, we replace the static $T_{\mathrm{c}}$ by a phenomenological time-dependent critical temperature
\begin{equation}
  T_{\mathrm{c}}^{*}(t)=T_{\mathrm{c}}\Big( 1 + H(t)\,\cdot s\,\cdot\,\exp\left(-t/ \tau_{\mathrm{ph\mbox{-}ph}}\right)\Big)\,,
   \label{equ:Tc(t)}
\end{equation}
with Heavyside step function $H$. It captures the enhanced critical temperature in the nonthermal regime, given by the temperature scaling $s$, and converges to $T_{\mathrm{c}}$ at late times. This leaves us with only two global fit parameters for the simulations: damping $\gamma$ and scaling $s$ in the nonthermal regime (see Supplementary Note\,\ref{supp:model} for details of the model). For the timescale connecting both regimes, we find a good description of the data by choosing the lattice thermalization time $\tau_{\mathrm{ph\mbox{-}ph}}= 2.2$\,ps reported for the closely related compound LaTe$_3$\supercite{dolgirev2020}. Energy redistribution processes within the electron and lattice systems are often modelled by a three temperature model (3TM)\supercite{perfetti2007, johnson2017}, as presented in Fig.\,\ref{fig:potentials}c. Here, $\tau_{\mathrm{ph\mbox{-}ph}}$ corresponds to the timescale of energy transfer between strongly coupled optical phonon modes ($T_{\mathrm{hot\mbox{-}ph}}$) with the remaining cold lattice modes ($T_{\mathrm{l}}$). The choice of the parameter $\tau_{\mathrm{ph\mbox{-}ph}}$ is further motivated in the following discussion. In this description, CDW order emerges when the electronic temperature $T_{\mathrm{e}}$ falls below the introduced dynamic effective $T_{\mathrm{c}}^{*}$ (black dashed curve in Fig.\,\ref{fig:potentials}c). During the thermalization process, the estimated lattice temperatures $T_\mathrm{l}$ stay below the thermal critical temperature for all applied fluences.

Given the complexity of the system, this model with its minimal amount of free parameters is in remarkable agreement with the electronic order parameter extracted directly from the trARPES data throughout the CDW melting and full recovery over a large fluence range, as shown in Figs.\,\ref{fig:shortdelays}f and \ref{fig:longtermrecovery}a. It captures (i) the AM in the low-fluence regime, (ii) the CDW melting time after arrival of the pump, (iii) the coherent oscillations and the down-chirp in the high-fluence regime, and (iv) the full CDW recovery to equilibrium. The fit yields a nonthermal critical temperature of $T_{\mathrm{c}}^{*}(t=0$\,fs$)\approx 745$\,K, i.e., more than double of the equilibrium $T_{\mathrm{c}}$. Remarkably, this value is similar to the electronic temperature where the onset of CDW recovery is observed in Fig.\,\ref{fig:longtermrecovery}b. To illustrate the necessity of a transiently enhanced $T_{\mathrm{c}}^{*}$ to describe the data, we perform tdGL simulations keeping the critical temperature fixed at the equilibrium value, which, however, leads to a severe deviation from the experimental oscillations and CDW recovery, see Supplementary Fig.\,\ref{fig:supp_sim}. Next, we illustrate the characteristic regimes of the tdGL simulations based on the extracted transient potential energy surfaces $V(\psi,t)$ in Fig.\,\ref{fig:potentials}.

\textbf{AM regime:} Before excitation, the system is in the CDW ground state ($\eta\approx0$), corresponding to an underlying double-well potential with minima at $|\psi|\approx 1$. Upon weak excitation (Fig.\,\ref{fig:potentials}a), the potential surface is barely altered and maintains its double-well shape. This launches a damped oscillation of the order parameter around the marginally shifted potential minimum at frequency $\omega_{\mathrm{AM}}$, i.e., the AM. 

\textbf{Overshoot regime:} Upon strong excitation (Fig.\,\ref{fig:potentials}b), the underlying potential transforms to a single-well shape, corresponding to the metallic phase. The order parameter overshoots to the opposite side of the potential, and oscillates around the new potential minimum at $|\psi| = 0$ at frequency $\omega \gg \omega_{\mathrm{AM}}$. Relaxation of the system leads to a transient flattening of the potential, resulting in the observed frequency down-chirp. At $\eta<1$, the CDW order finally recovers, and the order parameter relaxes into one of the minima of the emerging double-well potential.

A minor deviation of the fit from the data occurs at the dynamical slowing-down of the CDW melting in the vicinity of the melting threshold, as observed in the curve at fluence 0.05\,mJ cm$^{-2}$ in Fig.\,\ref{fig:shortdelays}f. For an initial perturbation in the range $\eta_{\mathrm{init}}\approx 0.5 \dots 1$, the system gains just enough energy to reach the local maximum of the double-well potential at $|\psi|=0$. Close to this metastable point, the potential is rather flat, leading to a critical slowing-down of the order-parameter dynamics\supercite{zong2019}, discussed in detail in Supplementary Note\,\ref{supp:slowdown}. A similar critical behaviour is expected during the recovery of the CDW order. In the overshoot regime, after dampening of the initial oscillations, the order parameter can get trapped at the metastable local maximum despite an incipient recovery of the double-well ground state. However, in real systems, several microscopic processes, such as local modification of $T_{\mathrm{c}}$ by crystal defects\supercite{arguello2014, fang2019}, CDW nucleation and creation of topological defects\supercite{zong2019evidence} and coupling of the collective excitation to other phonons\supercite{yusupov2008}, will screen against a pronounced critical slowing-down. However, such effects go beyond our current model.

To reproduce the main observations of the extracted structural order parameter, we extend this model to a layered description (see Supplementary Note\,\ref{supp:tdgl_structural}), as shown in Fig.\,\ref{fig:shortdelays}h. However, the absence of clear coherent modulations in the time evolution of the SL peak intensity and the additional contribution of the SL phase coherence prohibit a reliable fit of $I_{\mathrm{SL}}(t)$. Nonetheless, we conclude that this model captures all key features of the structural and electronic order parameters within a unified framework.

\section*{Discussion}
We unambiguously demonstrate a transient CDW behaviour distinct from equilibrium, as evidenced by the CDW AM modulations after weak excitation despite electronic temperatures exceeding thermal $T_\mathrm{c}$, and from the CDW recovery at elevated electronic temperatures after strong excitation. The qualitative correspondence of charge and structural features of the CDW excludes a scenario in which only the electronic superstructure is destroyed while the lattice distortion remains intact, which could facilitate such a nonthermal behaviour. So what causes this enhanced transient stability of CDW order far beyond the equilibrium $T_{\mathrm{c}}$? In equilibrium, lattice fluctuations induced by thermally populated phonons, accompanied by fluctuations of the charge density, reduce $T_{\mathrm{c}}$ significantly below the mean-field value $T_{\mathrm{MF}}$. Especially in low-dimensional systems, these fluctuation effects become increasingly important, such that long-range order and phase transitions cannot occur at finite temperatures in strictly 1D systems\supercite{motizuki1986, gruner1994}. However, in real materials, coupling between neighbouring chains stabilizes the CDW order, resulting in short-range correlations at high-temperatures and long-range 3D order below $T_{\mathrm{c}}$\supercite{gruner1994, pouget2016}.

\begin{figure}[!ht]
\centering
\includegraphics[]{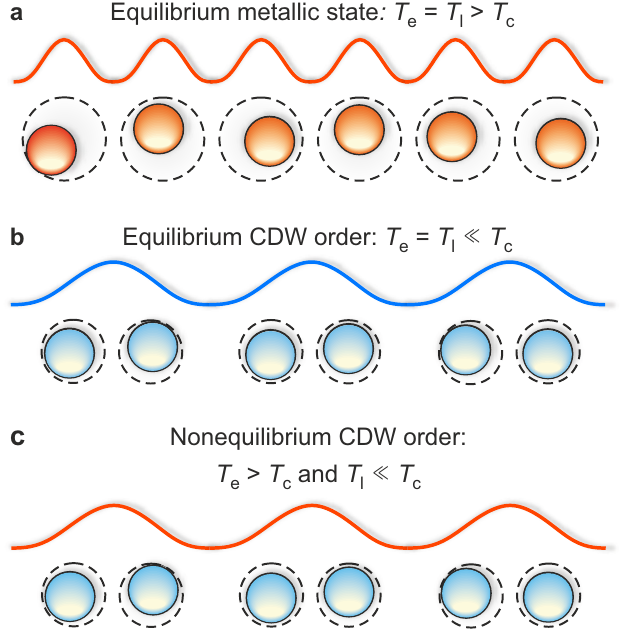}
\caption{\textbf{Illustration of nonthermal CDW order.} (\textbf{a}) In equilibrium at elevated temperatures, the system is in a trivial metallic phase. The charge density (wavy line) and the mean positions of the ionic cores (circles) are spaced evenly, as strong thermal lattice fluctuations prevent long-range CDW order. (\textbf{b}) In equilibrium at low temperatures, the system features an ordered charge- and lattice superstructure. (\textbf{c}) Photoexcitation of the CDW ground state ($T_{\mathrm{pre\mbox{-}exc.}}\ll T_{\mathrm{c}}$) generates a hot electron distribution, while the lattice initially remains cold. In this out-of-equilibrium state, thermal lattice fluctuations are weak and barely hinder long-range CDW ordering. Hence, the charge and lattice superstructure is stabilized at electronic temperatures beyond $T_{\mathrm{c}}$.}
\label{fig:fluctuations}
\end{figure}

Ultrafast optical perturbation breaks the thermal equilibrium between charges and lattice. Initially, electrons and certain optical phonons are strongly excited, while the overall vibrational population of the lattice -- determined by acoustic modes that account for the majority of the lattice heat capacity -- is still close to its pre-excitation value corresponding to an effective lattice temperature significantly below $T_{\mathrm{c}}$. In this out-of-equilibrium regime, the average displacement of the ionic cores around their mean positions (mean-squared displacement) is small, as the nonthermal phonon population is dominated by high-frequency, low-amplitude optical phonons\supercite{waldecker2016electron}. Thus, initially after excitation, lattice fluctuations are strongly suppressed and counteract a mean-field long-range ordering only weakly, which facilitates CDW formation even at electronic temperatures far beyond $T_{\mathrm{c}}$, illustrated in Fig.\,\ref{fig:fluctuations}. In this nonthermal regime, $T_{\mathrm{c}}$ is replaced by the effective electronic critical temperature $T_{\mathrm{c}}^{*}$, which is renormalized towards the mean field value depending on the transient lattice temperature and concomitant fluctuations. Over the course of several ps, depending on the lattice thermalization time $\tau_{\mathrm{ph\mbox{-}ph}}$, energy is transferred from the strongly coupled optical hot phonons to the remaining phonon modes. This defines the crossover from the nonthermal to the quasi-thermal regime, at which electrons and lattice locally reach thermal equilibrium. As the lattice temperature rises, acoustic (high-amplitude) fluctuations and CDW phase fluctuations increase, which impedes long-range 3D CDW order, and $T_{\mathrm{c}}^{*}$ consequently converges towards the equilibrium $T_{\mathrm{c}}$. The increasing occupation of lattice vibrations also increases the lattice entropy, and thus modifies the underlying free energy surface. In this picture, the changing lattice entropy plays the analogous part to the time-dependent critical temperature introduced within our tdGL expansion.

The agreement of the Ginzburg-Landau simulations with the extracted order parameters further underlines this scenario. The initial oscillation frequency of the electronic order parameter, the down-chirp as well as the recovery are reproduced by simulations with an enhanced $T_{\mathrm{c}}^{*}$, that converges towards the equilibrium $T_{\mathrm{c}}$ on the lattice thermalization time. In addition, since the initial lattice temperature is close to its equilibrium value also after strong excitation to the overshoot regime, the contribution of thermal fluctuations is expected to be rather independent of fluence. This is in agreement with our model, which captures the experimental data over a wide fluence range with a fluence-independent description of $T_{\mathrm{c}}^{*}$. Our simulations yield a transient critical temperature of $\approx 750$\,K at early times, which is still considerably below the mean field transition temperature $T_{\mathrm{MF}}\approx 1600$\,K estimated from the electronic energy gap in the nested regions via the well-known BCS expression\supercite{gruner1994}. However, because of the imperfect nesting of large segments of the FS, a significant reduction of $T_{\mathrm{MF}}$ is expected\supercite{yamaji1983, ru2008}, and remaining fluctuations at the initial lattice temperature of $T_{\mathrm{l}}\approx$100\,K are further expected to lead to a lower $T_{\mathrm{c}}^{*}$.

The CDW order above $T_{\mathrm{c}}$ may be further stabilized by transiently enhanced FS nesting. A previous trARPES study has demonstrated an improved nesting condition in rare-earth tritellurides upon optical excitation\supercite{rettig2016}, caused by a transient modification of the FS. Consequently, the CDW-gapped area at $E_{\mathrm{F}}$ expands and the critical temperature transiently increases. However, the photoinduced enhanced nesting significantly increases with excitation density\supercite{rettig2016}, which would result in a strongly fluence-dependent nonthermal critical temperature. As we find a good description of the data by $T_\mathrm{c}^*$ independent of fluence, we assign a suppression of lattice fluctuations in the out-of-equilibrium state as the dominant effect stabilizing the transient CDW. Several studies suggest similar nonthermal behaviour in other CDW materials. The commensurate CDW phase of 1T-TaS$_2$ exhibits an exceptionally robust AM after strong perturbation, with initial electronic temperatures exceeding 1300\,K\supercite{perfetti2006}. In elemental Chromium, trXRD measurements of the SL peak indicate a persisting CDW state above the thermal transition temperature\supercite{singer2016photoinduced}.

\section*{Conclusion}
In summary, we experimentally track the structural and electronic order parameters of a photoinduced CDW-to-metal transition in the rare-earth tritelluride TbTe$_3$, and reveal a close correspondence of the charge and lattice components of the CDW phase throughout the melting and initial recovery of order. By extracting the time-dependent electronic temperature, we demonstrated nonthermal CDW formation at electronic temperatures significantly above the thermodynamic transition temperature $T_{\mathrm{c}}$. We attribute the dominating role of this behaviour to reduced lattice fluctuations compared to a scenario in which charges and lattice are in equilibrium above $T_{\mathrm{c}}$. Since lattice fluctuations play a universal role in the CDW formation, the observed nonthermal stabilization mechanism should also apply to other material families. Moreover, we observed excitation-dependent collective dynamics of the charge order, closely connected to a coherent modulation of the periodic lattice distortion. We applied a tdGL framework to model the order-parameter dynamics and to describe the underlying transient potential energy surface, which governs the collective behaviour. Despite its simplicity of using a single degree of freedom, this phenomenological model reproduces all key observations. This suggests that mode-coupling\supercite{yusupov2008} and inhomogeneities (defects) play only a secondary role in the dynamical melting and recovery of the CDW amplitude.

As any memory device relies on nonequilibrium properties, our results have strong implications for applications involving charge-ordering phenomena. A key parameter defining the persistence of the nonthermal stabilization is phonon-phonon coupling, as it dictates the lattice thermalization and thus the timescale on which the fluctuation background rises. Therefore, minimizing phonon-phonon coupling may be critical in the design of switchable CDW devices operating in nonequilibrium conditions\supercite{vaskivskyi2015controlling}.

\section*{Methods}
\textbf{trARPES.} Single crystals of TbTe$_3$ samples were grown by slow cooling of a binary melt\supercite{ru2006}. All experiments were carried out at $T=100$\,K. The ARPES measurements were performed in ultra-high vacuum $<1$ $\times$ 10$^{-10}$\,mbar (samples cleaved in-situ), using a laser-based higher-harmonic-generation trARPES setup\supercite{puppin2019} ($h\nu_{\mathrm{probe}}$=21.7\,eV, $h\nu_{\mathrm{pump}}$=1.55\,eV, 500\,kHz repetition\@ rate, $\Delta E\approx$ 175\,meV, $\Delta t\approx$ 35\,fs) and a SPECS Phoibos 150 hemispherical analyzer. The pump and probe spot sizes (FWHM) are $\approx 230 \times 200$\,\textmu m$^2$ and $\approx 70  \times 60$\,\textmu m$^2$. All discussed fluence values refer to the absorbed fluence $F_\mathrm{abs}$. To estimate $F_\mathrm{abs}$, the complex refractive index was determined via optical reflectivity measurements at $\lambda$=800\,nm to $n$=0.9 and $k$=2.6.

\textbf{trXRD.} The trXRD measurements were carried out at the FEMTO hard X-ray slicing source (X05LA) at the Swiss Light Source, Paul Scherrer Institut, Villigen, Switzerland\supercite{ingold2007}. The utilized laser-sliced X-ray pulses ($h\nu_{\mathrm{X\mbox{-}ray}}$=7\,keV, $\Delta t\approx$ 120\,fs) feature the high stability of conventional synchrotron radiation and do not exhibit any relevant jitter in position, angle or wavelength. The diffracted X-ray intensity was recorded with an avalanche photodiode in an asymmetric diffraction geometry. A synchronized optical pump laser (10° angle of incidence, $h\nu_{\mathrm{pump}}$=1.55\,eV, $\Delta t\approx$ 110\,fs) was used to excite the system. The pump and probe spot sizes (FWHM) were $\approx 600 \times 600$\,\textmu m$^2$ and $\approx 250 \times 5$\,\textmu m$^2$. The X-ray extinction length was matched to the pump penetration depth of 25\,nm by using a grazing angle of incidence of 0.5°.

\section*{Acknowledgements}
We thank E.M.\ Bothschafter for support during the trXRD experiments. This work was funded by the Max Planck Society, the European Research Council (ERC) under the European Union's Horizon 2020 research and innovation program (Grant No. ERC-2015-CoG-682843), the German Research Foundation (DFG) within the Emmy Noether program (Grant No. RE 3977/1), and the DFG research unit FOR 1700. Crystal growth and characterization at Stanford University (P.W. and I.R.F.) was supported by the Department of Energy, Office of Basic Energy Sciences under Contract No. DE-AC02-76SF00515. Part of this work was supported by the NCCR Molecular Ultrafast Science and Technology (Grant No. 51NF40-183615), a research instrument of the Swiss National Science Foundation (SNSF).

\section*{Data availability}
The data that support the findings of this study are publicly available \red{in Zenodo with the identifier doi:10.5281/zenodo.4106272}\supercite{data2020}.

\section*{Author contributions}
Y.W.W, L.R., M.P., C.W.N. and J.M. carried out the trARPES experiments; L.R., V.E., M.P., J.R., D.L., M.K., M.S., E.A., S.L.J., P.B., G.I. and U.S. carried out the trXRD experiments; P.W. and I.R.F. provided the samples; J.M. analysed the data with support from L.R; J.M. wrote the manuscript with support from L.R., R.E. and M.W.; all authors commented on the paper.

\section*{Competing Interests}
\red{The authors declare no competing interests.}

\section*{Correspondence}
Correspondence should be addressed to \href{mailto:maklar@fhi-berlin.mpg.de}{\nolinkurl{maklar@fhi-berlin.mpg.de}} and \href{mailto:rettig@fhi-berlin.mpg.de}{\nolinkurl{rettig@fhi-berlin.mpg.de}}.

\newpage
\beginsupplement
\appendix
\section*{Supplementary Information for "Nonequilibrium Charge-Density-Wave Order Beyond the Thermal Limit"} 
\section{Supplementary Figures}
\label{supp:supp_figures}

\begin{figure*}[!ht]
\centering
\includegraphics[]{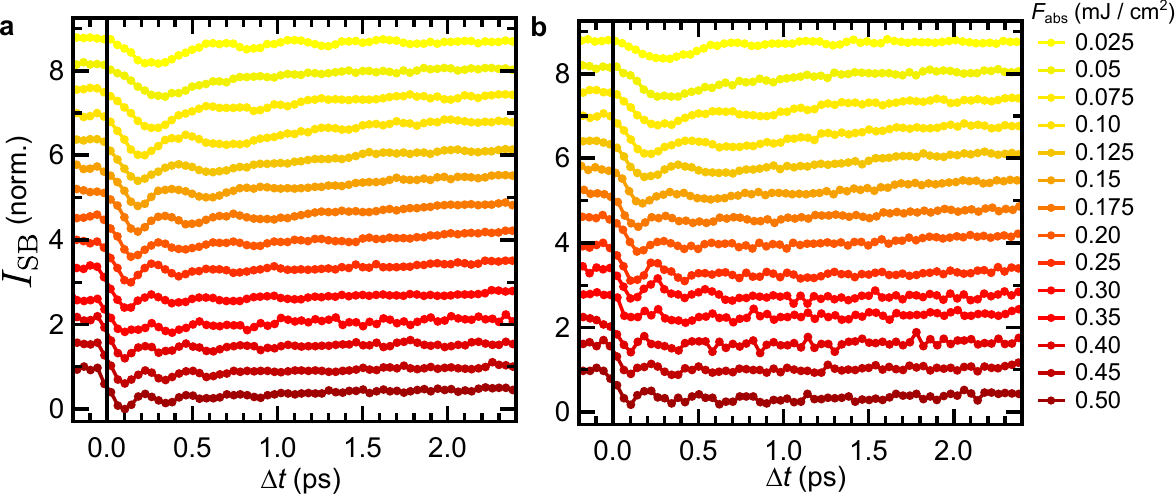}
\caption{\textbf{Shadow band intensity dynamics.} Shadow band intensities extracted from (\textbf{a}) the nested (gapped) region of the FS, see box 2 in Fig.\,\ref{fig:shortdelays}b, and (\textbf{b}) from the imperfectly nested (metallic) region, see box 3. Despite a slightly lower data quality in \textbf{b}, both shadow bands exhibit identical behaviour over the entire fluence range. The curves are vertically offset for clarity. For each curve, an intensity background extracted from a box slightly horizontally offset from the shadow band position is subtracted. Further, all curves are normalized by their respective intensities before excitation.}
\label{fig:SB_data}
\end{figure*}

\begin{figure*}[!ht]
\centering
\includegraphics[]{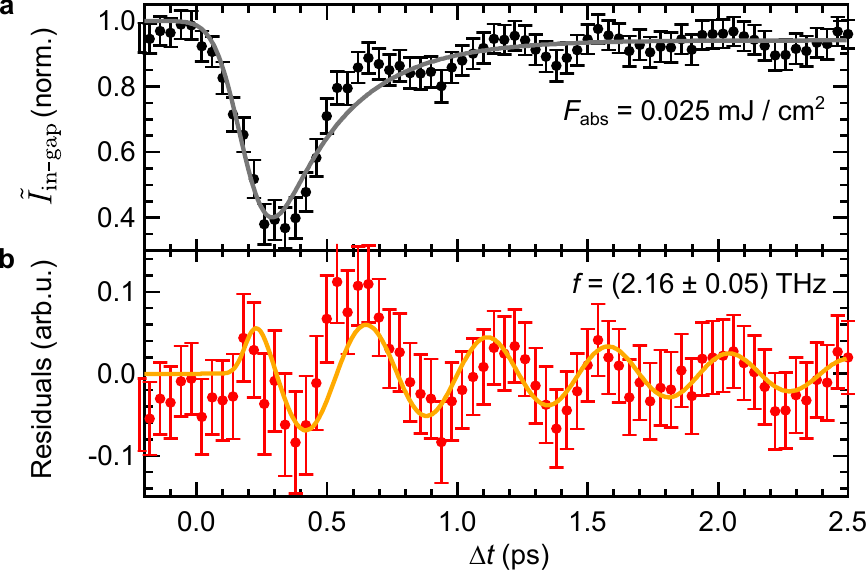}
\caption{\textbf{Electronic AM dynamics.} (\textbf{a}) Time evolution of the inverted in-gap intensity after weak excitation. The grey line marks a double-exponential fit convolved with a Gaussian. (\textbf{b}) Fit residuals showing a pronounced amplitude mode in agreement with previous trARPES experiments\supercite{schmitt2008, schmitt2011}, with a damped sinusoidal fit (orange curve). The error bars correspond to one standard deviation from electron counting statistics.}
\label{fig:supp_CDW_AM}
\end{figure*}

\begin{figure*}[!ht]
\centering
\includegraphics[]{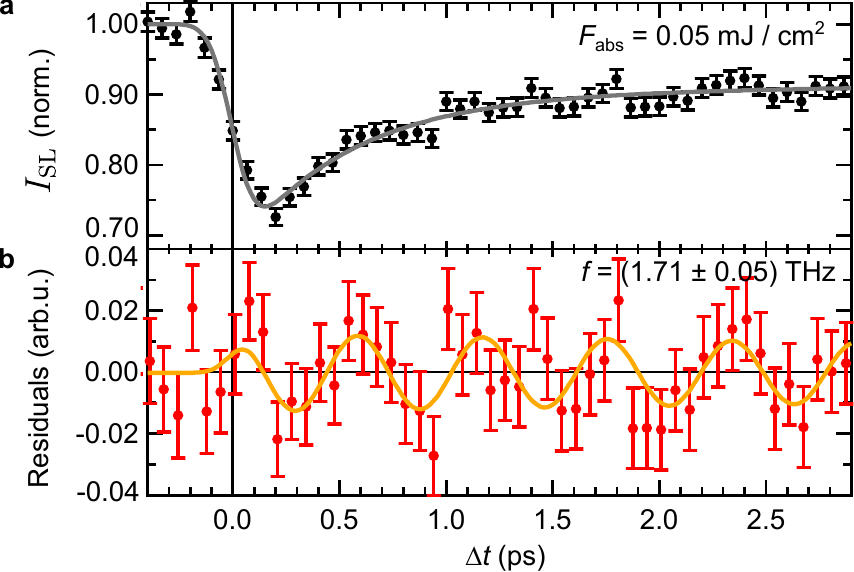}
\caption{\textbf{Structural AM dynamics.} (\textbf{a}) Time evolution of the normalized (2 10 1+$q_{\mathrm{CDW}}$) SL peak intensity after weak excitation. The grey line marks a double-exponential fit convolved with a Gaussian. \red{The error bars correspond to one standard deviation from photon counting statistics.} (\textbf{b}) The fit residuals indicate weak oscillations superimposed on the exponential decay, corresponding to a phonon mode that strongly couples to the CDW amplitude mode at 100\,K, in agreement with previous optical and trXRD studies\supercite{yusupov2008, moore2016, maschek2018competing, trigo2019}, with a damped sinusoidal fit (orange curve).}
\label{fig:supp_X_ray_AM}
\end{figure*}

\begin{figure*}[!ht]
\centering
\includegraphics[]{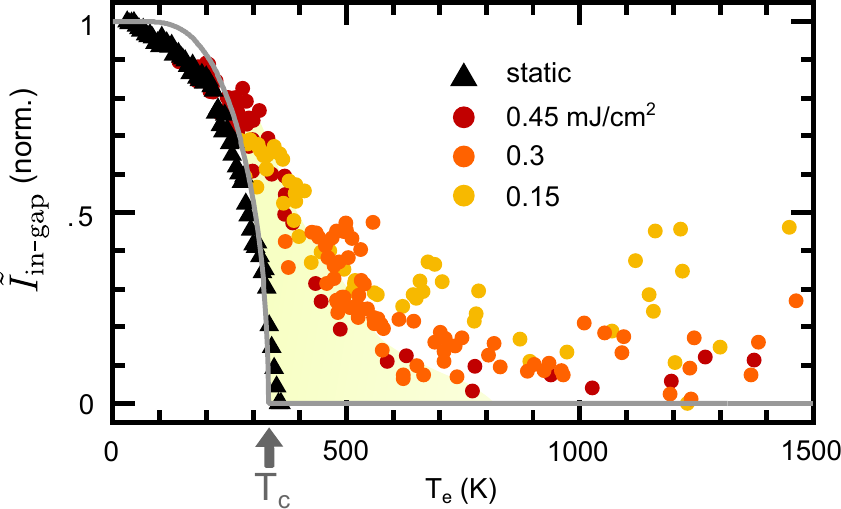}
\caption{\textbf{$T$-dependent CDW recovery dynamics.} Inverted in-gap intensity of the dynamic CDW recovery versus extracted electronic temperatures for selected fluences. For reference, the static $T$-dependence is shown (black) with the BCS-like mean-field curve (grey). For clarity, the values of the dynamic traces for $t<$200\,fs (initial CDW melting) are omitted. The dynamic curves follow a universal recovery behaviour over a wide range of fluences. The region of nonthermal CDW order above the thermal critical temperature is shaded yellow. Normalization of the dynamic traces according to Fig.\,\ref{fig:longtermrecovery}b.}
\label{fig:supp_ingap_Te}
\end{figure*}

\clearpage

\section{Details of the tdGL simulations of the electronic order parameter}
\label{supp:model}
To simulate the observed electron dynamics, we solve the following equation of motion based on the transient potential energy surface (Eq.\,\ref{equ:potential})

\begin{equation}
   \frac{\partial^2}{\partial t^2}\psi = \frac{\omega_{\mathrm{AM}}^2}{2} \Big(\big(1-\eta(t) \big) \psi - \psi^3\Big) - \gamma \frac{\partial}{\partial t}\psi \,,
   \label{equ:motion}
\end{equation} 

which yields the order parameter $\psi(t)$ used to simulate the diffracted intensities and the in-gap photoemission intensities. The initial conditions are chosen as 
\begin{equation*}
    \psi= \sqrt{1-\frac{T_{\mathrm{base}}}{T_{\mathrm{c}}}}\approx0.84\,,
\end{equation*}
i.e., the static Ginzburg-Landau value corresponding to the temperature before excitation, and 
\begin{equation*}
    \frac{\delta \psi}{\delta t}=0\,.
\end{equation*}
We perform a global fit of the electronic in-gap dynamics $\tilde{I}_{\mathrm{in\mbox{-}gap}}(t)$ over the full accessible fluence range (Fig.\,\ref{fig:shortdelays}f) with the free parameters damping $\gamma$ and scaling factor $s$ of the nonthermal critical temperature, see Eq.\,\ref{equ:Tc(t)}, while the remaining input parameters are fixed (\red{Supplementary} Table\,\ref{table:parameters}). In order to fit the inverted in-gap intensity, the order-parameter simulations are normalized. To model the transient potential energy surface, see Eq.\,\ref{equ:potential}, we use the extracted electronic temperatures in a parametrized form, see Supplementary Note\,\ref{supp:Te}. We find that the maximum electronic temperature yields a good description of the initial excited potential energy shape. This is evident from the saturation of $T_{\mathrm{e, max}}$ in the high-fluence regime (see Supplementary Fig.\,\ref{fig:Te_multi}c) that is accompanied by an upper limit of the initial coherent modulation frequency of the electronic order parameter. This also implies that the potential energy surface does not directly scale with the absorbed fluence $\eta\not\propto F$. 

We aim to define the fit parameters as simple as possible; however, we can not reproduce the experimental data over the entire fluence- and temporal range with a single global damping constant. While $\gamma$ correctly captures the initial damped modulations, a constant damping results in the reappearance of coherent oscillations in the high-fluence regime, when the potential transforms back from the high-symmetry to the double-well shape. To prevent this, we use an alternative global fit parameter $\gamma_{\mathrm{rec}}$ during the recovery ($>2$\,ps) in the high-fluence regime ($\geq 0.3$\,mJ cm$^{-2}$). In real systems, dephasing prevents the reappearance of coherent oscillations during the recovery.

We account for the inhomogeneous excitation profile, corresponding to the pump and probe spot sizes (FWHM) of $\approx 230 \times 200$\,\textmu m$^2$ and $\approx 70 \times 60$\,\textmu m$^2$, respectively, by averaging over multiple simulations with varying fluences (up to $\pm 7.5$\,\% around the centre value). Finally, to account for the temporal resolution of the experiment, the simulations are convolved with a Gaussian (FWHM $=35$\,fs).\\

\renewcommand{\arraystretch}{1.3}
\begin{table*}[!ht]
\caption{Parameters of the tdGL simulations\label{table:parameters}}
\begin{center}
 \begin{tabular}{|c |c |c| } 
 \hline
 Parameter & Value & Physical meaning \\ [0.5ex] 
 \hline\hline
 $\omega_{\mathrm{AM}} / 2\pi$ & 2.2 THz\supercite{yusupov2008} & AM at 100\,K \\ 
 \hline
 $\gamma$ & 4.4 THz & Damping \\ 
 \hline
 $\gamma_{\mathrm{rec}}$ & 11.3 THz & Damping during the recovery $> 2\,$ps in the overshoot regime \\ 
 \hline
 $T_\mathrm{c}$ & 336 K\supercite{ru2008} & Critical temperature of the CDW \\
 \hline
 $s$ & 1.22 & Scaling factor of the rescaled critical temperature $T_{\mathrm{c}}^{*}$ \\
 \hline
 $\tau_{\mathrm{ph\mbox{-}ph}}$ & 2.2\,ps\supercite{dolgirev2020} & Decay constant of the rescaled critical temperature $T_{\mathrm{c}}^{*}$ \\
 \hline
\end{tabular}
\end{center}
\end{table*}

\begin{figure*}[!ht]
\centering
\includegraphics[]{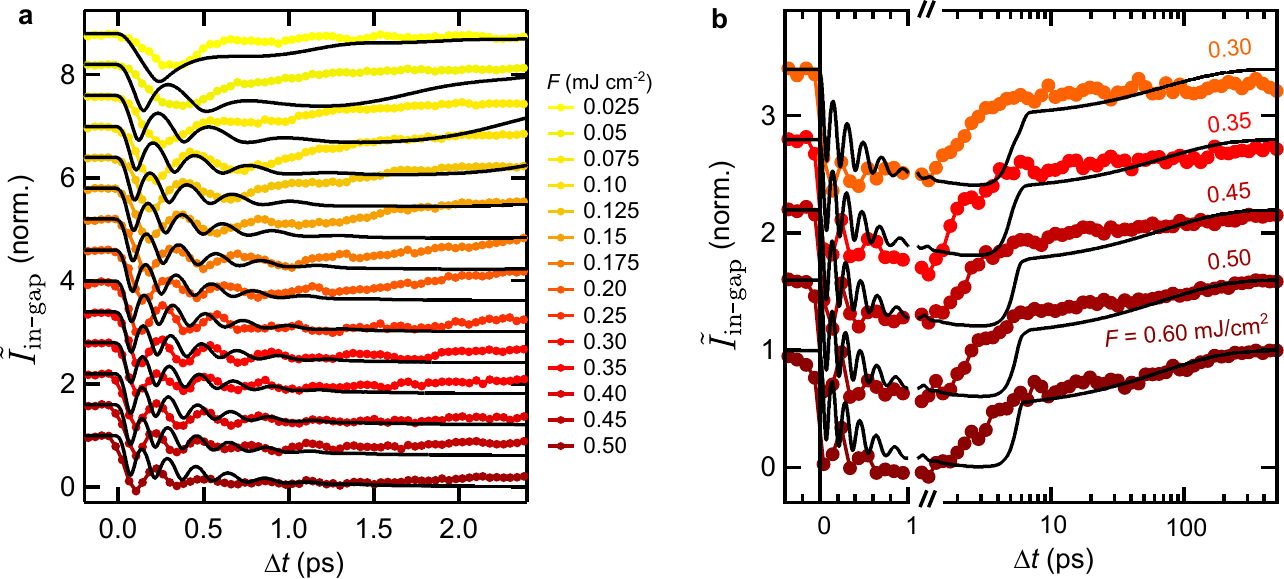}
\caption{\textbf{tdGL simulations using the constant equilibrium critical temperature $T_\mathrm{c}$.} (\textbf{a}) Experimental data analog to Fig.\,\ref{fig:shortdelays}f and (\textbf{b}) to Fig.\,\ref{fig:longtermrecovery}a of the main manuscript. The tdGL simulations are performed using the parameters as described above, however, using a fixed critical temperature of $T_\mathrm{c}=336$\,K.}
\label{fig:supp_sim}
\end{figure*}

As illustrated in Fig.\,\ref{fig:longtermrecovery}b, the transient CDW recovery strongly deviates from static (thermal) behaviour, which necessitates the introduction of a transiently increased critical temperature $T_c^*$ in the tdGL simulations. To highlight the requirement of a transiently increased $T_c^*$, we perform additional simulations employing the constant equilibrium critical temperature $T_\mathrm{c}$, while keeping the remaining parameters fixed as described above. As Supplementary Fig.\,\ref{fig:supp_sim} shows, this does not reproduce the experimental data, as (i) the simulated oscillation frequencies are strongly overestimated due to the increased slope of the underlying potential energy landscape (determined by $\eta(t)$) and (ii) the simulated recovery sets in only at $T_\mathrm{e}<T_c$ -- at a significant delay with respect to the experimental data.

\section{Parametrization of the electronic temperatures}
\label{supp:Te}
In the tdGL simulations, the electronic temperature $T_\mathrm{e}(t)$ enters as an input parameter that determines the underlying potential shape. Thus, we extract the transient electronic temperatures from Fermi-Dirac fits of the quasiparticle energy distribution of the metallic region of the FS\supercite{wang2012, dolgirev2020}, as these values are more reliable than the approximation by the 3TM. For each dataset, the energy resolution ($\Delta E\approx 175$\, meV) is determined from a fit to energy distribution curves (EDCs) before the arrival of the pump pulse, fixing the base temperature to $T_{\mathrm{base}}=100$\,K. Then, the electronic temperature is extracted for varying delays, keeping the energy resolution fixed while using the position of the Fermi level and temperature as free fit parameters. Exemplary fits are shown in Supplementary Fig.\,\ref{fig:Te_single}a. A deviation from a thermal distribution appears for EDCs close to temporal pump-probe overlap, resulting in a large standard deviation of the extracted temperatures. Figure\,\ref{fig:Te_single}b depicts the electronic temperature evolution in the high-fluence regime. The relaxation of $T_{\mathrm{e}}$ features two distinct timescales, which we assign to the initial energy transfer from the electrons to specific optical phonons and a subsequent cooling of the thermalized system via diffusion.

\begin{figure*}[!ht]
\centering
\includegraphics[]{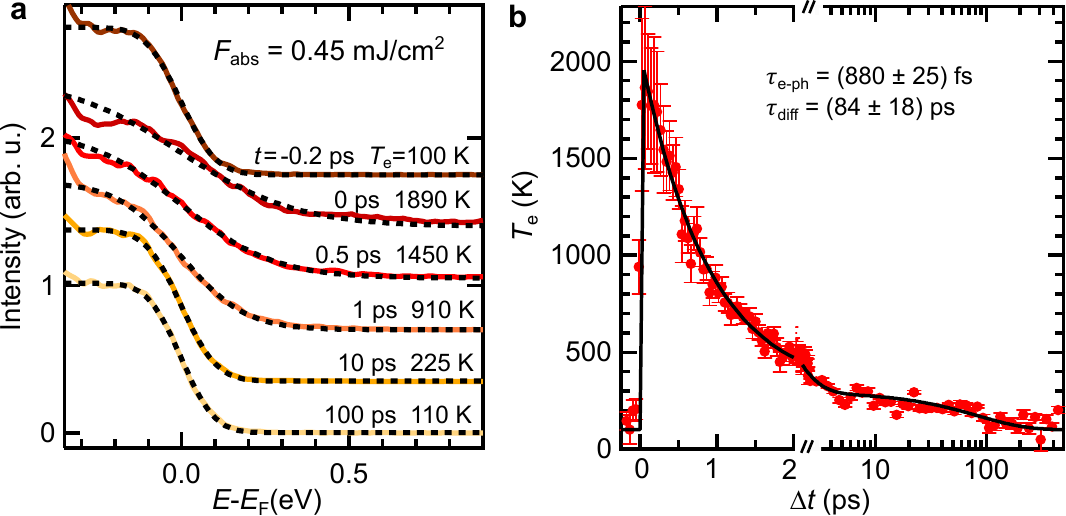}
\caption{\textbf{Time-dependent Fermi-Dirac fits}. (\textbf{a}) EDCs extracted from the metallic region of the FS with Fermi-Dirac fits for selected pump-probe delays. (\textbf{b}) Extracted electronic temperature as function of delay with biexponential decay fit (black curve). One standard deviation of the temperature fits are given as uncertainty.}
\label{fig:Te_single}
\end{figure*}

This fitting routine has been performed for all measured fluences in order to parametrize the fluence and time dependence of $T_{\mathrm{e}}$, shown in Supplementary Fig.\,\ref{fig:Te_multi}. The temporal evolution is approximated by a double-exponential decay:
\begin{equation}
   T_{\mathrm{e}}(t, F)=T_{\mathrm{base}}+H(t) \cdot T_{\mathrm{sat}}(F) \, \Big[ A_0 \cdot \exp (-t/\tau_{\mathrm{e\mbox{-}ph}}) + (1-A_0) \cdot \exp (-t/\tau_{\mathrm{diff}}) \Big]
   \label{equ:Te_t}
\end{equation}
with Heaviside step function $H(t)$, the excitation-dependent temperature increase $T_{\mathrm{sat}}$ discussed below, and the amplitude ratio between the fast ($\tau_{\mathrm{e\mbox{-}ph}}$) and slow ($\tau_{\mathrm{diff}}$) decay components. The values of the temperature parametrization are listed in \red{Supplementary} Table\,\ref{table:parameters_Te}. 

\begin{figure*}[!ht]
\centering
\includegraphics[]{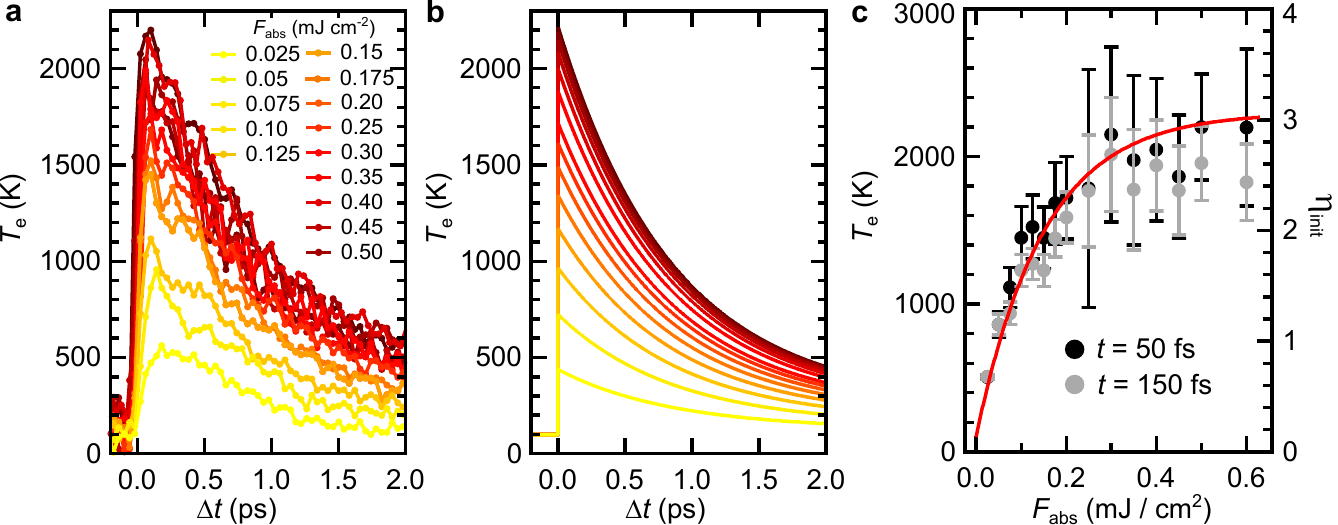}
\caption{\textbf{Electronic temperature parametrization}. (\textbf{a}) Extracted temporal evolution of electronic temperatures and (\textbf{b}) parametrization by \red{Supplementary} Eq.\,\ref{equ:Te_t}. (c) Extracted electronic temperatures close to temporal pump-probe overlap versus fluence. The saturation model of the maximum electronic temperatures (\red{Supplementary} Eq.\,\ref{equ:Te_max}) is shown in red. \red{One standard deviation of the temperature fits are given as uncertainty.}}
\label{fig:Te_multi}
\end{figure*}

In the regime of strong excitation, the maximum electronic temperatures saturate at $T_{\mathrm{e, max}}(t\approx 0\,$fs$)\approx 2300$\,K (see Supplementary Fig.\,\ref{fig:Te_multi}c). While the electronic system has not fully thermalized close to pump-probe overlap (and therefore electronic temperatures are ill-defined), we find that this saturation trend is also evident at later pump-probe delays. This saturation effect can be either due to a highly nonlinear electronic heat capacity or due to photobleaching. As the FS of TbTe$_3$ consists of metallic and CDW-gapped regions, the electronic heat capacity is expected to follow a linear metal-like temperature dependence with an additional nonlinear increase resulting from the redistribution of spectral weight due to the phase transition\supercite{lin2008, dolgirev2020}. Furthermore, our observations agree with the saturation plateau of excited quasiparticle intensity in the related compound LaTe$_3$\supercite{zong2019}. Such a fluence saturation trend of the electronic excitation level has also been observed in Blue Bronze\supercite{neugebauer2019}.

We model the fluence-dependence of the temperature saturation as
\begin{equation}
  T_{\mathrm{sat}}(F)=T_0 \cdot \left[1-\exp(-F/f)\right], 
   \label{equ:Te_F}
\end{equation}
with the upper temperature limit $T_0$ and the fluence scaling factor $f$. The maximum electronic temperature is therefore given by
\begin{equation}
T_{\mathrm{e, max}}(F)=T_{\mathrm{base}}+T_{\mathrm{sat}}\,.
\label{equ:Te_max}
\end{equation}

\begin{table*}[!ht]
\caption{Parametrization values of the electronic temperature.\label{table:parameters_Te}}
\begin{center}
 \begin{tabular}{|c |c |c| } 

 \hline
 Parameter & Value & Physical meaning \\ [0.5ex] 
    \hline\hline
   $T_{\mathrm{base}}$ & 100\,K & Base temperature of the sample\\
 \hline
 $T_0$ & 2200\,K & Temperature limit of the saturation model \\ 
 \hline
 $f$ & 0.15\,mJ cm$^{-2}$ & Fluence scaling factor \\ 
 \hline
 $\tau_{\mathrm{e\mbox{-}ph}}$ & 0.85\,ps & Fast decay constant of the electronic temperature evolution \\  \hline
 $\tau_{\mathrm{diff}}$ & 85\,ps & Slow decay constant of the electronic temperature evolution \\ 
 \hline
 $A_0$ & 0.92 & Amplitude ratio between the two components of the biexponential decay\\
  \hline
\end{tabular}
\end{center}
 \label{table:Te}
\end{table*}

\section{tdGL simulations of the structural order parameter}
\label{supp:tdgl_structural}
To simulate the trXRD measurements of the SL peak intensity, we have to account for the contribution of sub-surface layers of varying excitation densities due to the finite pump and probe beam penetration depths. We introduce a layered model, in which the tdGL equation of motion is solved for each individual layer. The diffracted X-ray beam corresponding to the (2 10 1+$q_{\mathrm{CDW}}$) SL reflection leaves the sample at an exit angle of $\theta \approx$\,35°. As the lateral CDW correlation length $L_\mathrm{coh}$\supercite{ru2008} is significantly larger than the effective penetration depth of the X-ray field $L_\mathrm{coh} \gg 2\delta_{\mathrm{X\mbox{-}ray}}/\sin{\theta}$, interference of different layers has to be considered\supercite{beaud2014, rettig2016itinerant}. Thus, the total intensity is given by the coherent sum of all layers $j$ of thickness $d$  
\begin{equation}
    I_{\mathrm{SL}}(t) \propto \left( \sum_{j=0}^{\infty}  \exp{(-jd/2\delta_{\mathrm{X\mbox{-}ray}})}\cdot \psi_j(t)\right)^2 \,,
\end{equation}
whereas the first term weights the contribution of each layer according to the X-ray penetration depth $\delta_{\mathrm{X\mbox{-}ray}}=25$\,nm. The initial excitation level of the first layer $\eta_{0,\mathrm{init}}$ is calculated from the fluence-to-electronic-temperature calibration obtained from the trARPES data (see Supplementary Note \ref{supp:Te}). The attenuation of the excitation of burried layers is given by Lambert-Beer's law $\eta_{j,\mathrm{init}}=\eta_{0, \mathrm{init}} \cdot \exp(-jd/\delta_{\mathrm{pump}})$\,, with the penetration depth of the optical pulses $\delta_{\mathrm{pump}}=25$\,nm. We choose a layer thickness of $d=1$\,nm and sum the 250 topmost layers. To account for the temporal resolution of the experimental setup, the simulated intensity is convolved with a Gaussian (FWHM of 160\,fs). In the regime of very weak excitation, the introduced model leads to artifacts, as the rescaling of $T_\mathrm{c}^*$ causes an initial increase of $\psi(t)$ in cases where the electronic temperature barely increases. To avoid these simulation artifacts from buried layers at very low excitation densities, the order parameter $\psi_j(t)$ of layers $j$ with excitation levels $\eta_\mathrm{j,init}<0.25$ is fixed at the pre-excitation value $\psi_j(t<0)$.

As we do not observe clear oscillations of $I_\mathrm{SL}$ upon strong excitation, we do not include the trXRD data in the global fitting procedure. Rather, we apply the parameters of the simulations of the electronic order parameter to the layered model. In agreement with previous studies, we find that the dominant oscillatory component of the SL peak intensity after weak excitation is a $\approx1.7$\,THz mode\supercite{moore2016, trigo2019}. As the AM softens upon cooling, it crosses the energy of this additional mode, leading to an anti-crossing behaviour. Due to their coupling, this phonon mode appears at the same wave vector as the CDW\supercite{yusupov2008, maschek2018competing}, see Supplementary Fig.\,\ref{fig:supp_X_ray_AM}. Thus, we use $\omega_{\mathrm{AM}}=1.7$\,THz to simulate $|\psi_\mathrm{s}|$. Further, we omit the averaging over varying fluences, used in the simulations of the electronic order parameter. The remaining parameters are adopted from Supplementary Note\,\ref{supp:model}.

As discussed in the main text, surface steps may lead to unexcited sample areas. Therefore, a SL background persists even after strong excitation. To account for this, we rescale all structural intensity simulations by a global factor according to the maximum suppression of $I_\mathrm{SL}$ at the highest fluences.

This layered model captures all main experimental features of $I_\mathrm{SL}$, see Fig.\,\ref{fig:shortdelays}h. The absence of the oscillatory component in the high-fluence regime is well reproduced by the simulations, and results from the limited temporal resolution and the superposition of layers with varying excitation densities. Further, the absence of a recovery after strong excitation for several ps is in agreement with the simulations, and results from a destructive interference of the contributions of different layers with opposite sign of $\psi$, corresponding to opposite sides of the underlying potential\supercite{trigo2020formation}. In the low-fluence regime, the absolute intensities slightly deviate from the simulations. The absorbed fluence (determining the initial electronic temperature) is a highly sensitive input parameter of this model. Minor deviations between the fluence calibration of the trARPES and the trXRD setup have a major impact on the simulations. In addition, small uncertainties of the angle of incidence of the X-ray beam affect the penetration depth, a further sensitive parameter of this model. 

\section{Critical slowing-down of the CDW melting and recovery}
\label{supp:slowdown}

Critical slowing-down is a ubiquitous signature of phase transitions close to equilibrium\supercite{goldenfeld1992}, and can also occur in a dynamical setting upon perturbation\supercite{dolgirev2020self}. For several CDW systems, a dynamical slowing-down of the CDW melting after optical excitation in the regime of the threshold fluence has been observed\supercite{tomeljak2009, zong2019}. Here, we present a further instance of a dynamical slowing-down, which we discuss within the tdGL framework.

First, we utilize the tdGL formalism to study $t_{\mathrm{melt}}$, i.e., the CDW melting time, as function of excitation density, shown in Supplementary Fig.\,\ref{fig:slowdown_melting}b. For clarity, we turn off the relaxation of the potential energy surface after excitation ($\eta=\text{const}$), fix the critical temperature $T_{\mathrm{c}}^*$, and suppress damping. In the regime of weak perturbation, the first minimum appears at half the period of the AM. With increasing fluence, the CDW melting time increases and finally diverges at $\eta=0.5$. In the divergent case, see Supplementary Fig.\,\ref{fig:slowdown_melting}a, the energy gain of the excitation is just enough so that the order parameter approaches the local maximum $|\psi|\approx 0$. Close to this metastable point, the potential energy surface is fairly flat, which leads to a dynamical slowing-down. However, when using realistic simulation parameters, such as a relaxing potential energy surface, and taking into account an inhomogeneous excitation profile, the divergence is strongly reduced, and the simulated melting time agrees with the experimental data (see Supplementary Fig.\,\ref{fig:slowdown_melting}c).

\begin{figure*}[!ht]
\centering
\includegraphics[width=1\textwidth]{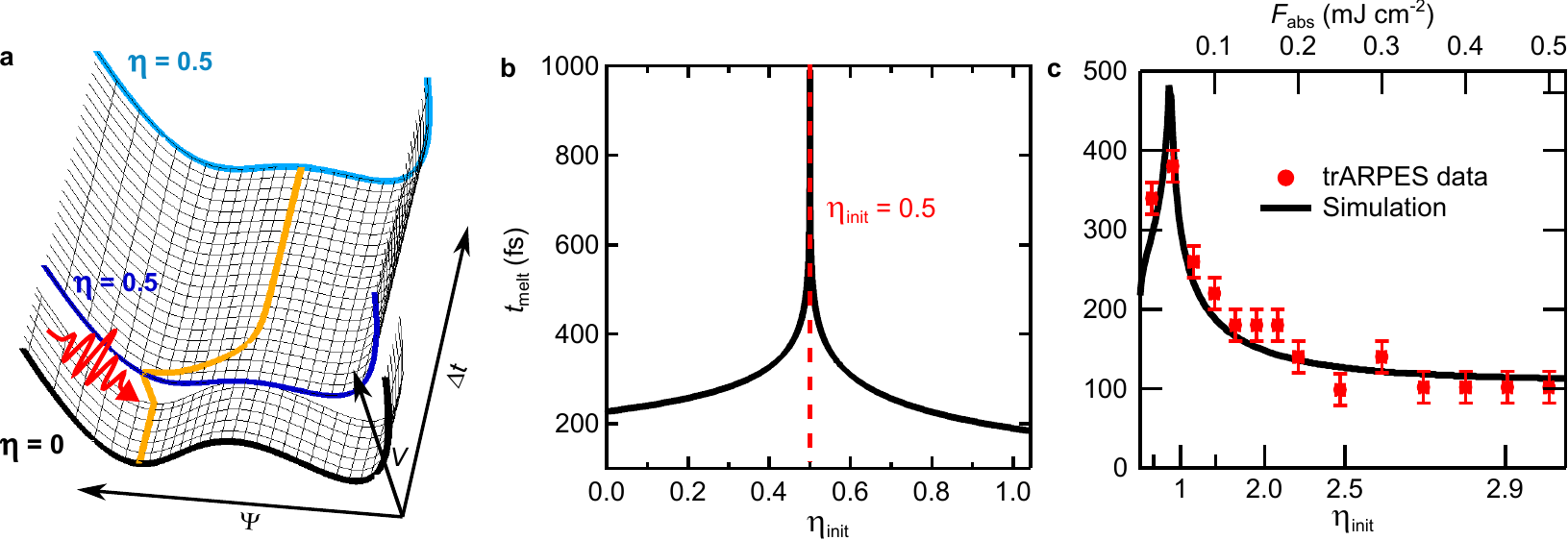}
\caption{\textbf{Dynamic slowing-down of the CDW melting}. (\textbf{a}) Transient potential and simulated order parameter (orange line) upon excitation corresponding to $\eta=0.5$. (\textbf{b}) Time to reach the first local minimum as function of initial excitation $\eta_{\mathrm{init}}$. Simulation parameters of a and b: $\gamma=0$\,THz, $T_{\mathrm{c}}^*=\text{const}=336$\,K  and $\tau_{\mathrm{e\mbox{-}ph}}=\tau_{\mathrm{diff}}=\infty$. (\textbf{c}) Initial minima of the inverted in-gap intensity, see Fig.\,\ref{fig:shortdelays}f, versus absorbed fluence and initial excitation $\eta_{\mathrm{init}}$. Results of the tdGL simulations with realistic model parameters (see Supplementary Note\,\ref{supp:model}) are shown in black. The error bars of the experimentally extracted melting times represent the temporal width (FWHM) of the XUV probe pulses.}
\label{fig:slowdown_melting}
\end{figure*}

A further dynamical slowing-down can occur during the recovery of the CDW. At specific fluences, when $|\psi|\approx 0$ and $\delta \psi / \delta t \approx 0$ at the same time as the potential regains the double-well shape ($\eta=1$), the order parameter gets frozen, illustrated in Supplementary Fig.\,\ref{fig:slowdown_recovery}. Due to the weak curvature in the vicinity of $|\psi|= 0$, the system is trapped in a metallic phase, despite an emerging double-well potential. However, this divergence is difficult to observe experimentally, as it occurs at narrow fluence windows, and is, similar to the slowing-down of the CDW melting, suppressed by crystal defects, coupling to other phonon modes and an inhomogeneous excitation profile. This critical behaviour leads to a delayed onset of CDW recovery in the simulations as compared to the electronic CDW dynamics for certain fluences, see Fig.\,\ref{fig:shortdelays}f (curve $F=0.1$\,mJ cm$^{-2}$) and Fig.\,\ref{fig:longtermrecovery}a.

\begin{figure*}[!ht]
\centering
\includegraphics[]{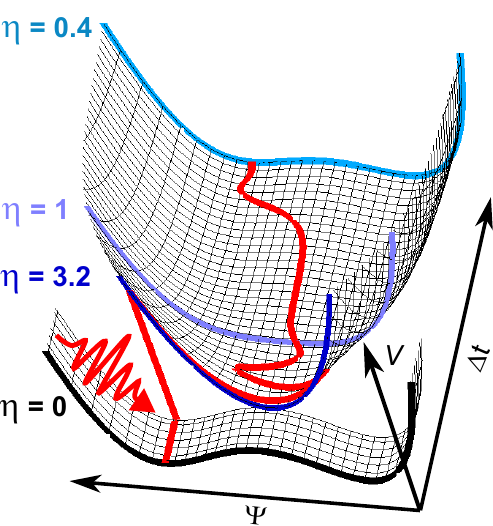}
\caption{\textbf{Dynamic slowing-down of the CDW recovery}. Transient potential and simulated order-parameter pathway of the dynamical slowing-down during the CDW recovery. For specific excitation conditions, the order-parameter dynamics critically slow down during recovery of the CDW double-well potential. Despite the appearance of the double-well shape for $\eta<1$ (purple line), the order parameter can get trapped at the metastable point $|\psi|\approx 0$, before it relaxes into one of the global minima. Model parameters are chosen analogous to Supplementary Note\,\ref{supp:model}. To demonstrate a pronounced slowing-down, the averaging over multiple curves with varying fluences is omitted.}
\label{fig:slowdown_recovery}
\end{figure*}

\clearpage

\printbibliography

@article{voit2000,
  title={Electronic structure of solids with competing periodic potentials},
  author={Voit, J and Perfetti, L and Zwick, F and Berger, H and Margaritondo, G and Gr{\"u}ner, G and H{\"o}chst, H and Grioni, M},
  journal={Science},
  volume={290},
  number={5491},
  pages={501--503},
  year={2000},
  publisher={American Association for the Advancement of Science}
}

@article{puppin2019,
  title={Time-and angle-resolved photoemission spectroscopy of solids in the extreme ultraviolet at 500 k{H}z repetition rate},
  author={Puppin, Michele and Deng, Yunpei and Nicholson, CW and Feldl, Johannes and Schr{\"o}ter, NBM and Vita, Hendrik and Kirchmann, PS and Monney, Claude and Rettig, Laurenz and Wolf, Martin and others},
  journal={Review of Scientific Instruments},
  volume={90},
  number={2},
  pages={023104},
  year={2019},
  publisher={AIP Publishing}
}

@article{ingold2007,
  title={Technical Report: FEMTO: A Sub-ps Tunable Hard X-ray Undulator Source for Laser/X-ray Pump-Probe Experiments at the SLS},
  author={Ingold, G and Beaud, P and Johnson, SL and Grolimund, D and Schlott, V and Schmidt, Th and Streun, A},
  journal={Synchrotron Radiation News},
  volume={20},
  number={5},
  pages={35--39},
  year={2007},
  publisher={Taylor \& Francis}
}

@article{rettig2014,
  title={Coherent dynamics of the charge density wave gap in tritellurides},
  author={Rettig, L and Chu, J-H and Fisher, IR and Bovensiepen, Uwe and Wolf, Martin},
  journal={Faraday discussions},
  volume={171},
  pages={299--310},
  year={2014},
  publisher={Royal Society of Chemistry}
}

@article{dolgirev2020,
  title={Amplitude dynamics of the charge density wave in {LaTe$_3$}: Theoretical description of pump-probe experiments},
  author={Dolgirev, Pavel E and Rozhkov, AV and Zong, Alfred and Kogar, Anshul and Gedik, Nuh and Fine, Boris V},
  journal={Physical Review B},
  volume={101},
  number={5},
  pages={054203},
  year={2020},
  publisher={APS}
}

@article{zong2019,
  title={Dynamical slowing-down in an ultrafast photoinduced phase transition},
  author={Zong, Alfred and Dolgirev, Pavel E and Kogar, Anshul and Erge{\c{c}}en, Emre and Yilmaz, Mehmet B and Bie, Ya-Qing and Rohwer, Timm and Tung, I-Cheng and Straquadine, Joshua and Wang, Xirui and others},
  journal={Physical review letters},
  volume={123},
  number={9},
  pages={097601},
  year={2019},
  publisher={APS}
}

@article{wang2012,
  title={Measurement of intrinsic Dirac fermion cooling on the surface of the topological insulator {Bi$_2$Se$_3$} using time-resolved and angle-resolved photoemission spectroscopy},
  author={Wang, YH and Hsieh, D and Sie, EJ and Steinberg, H and Gardner, DR and Lee, YS and Jarillo-Herrero, P and Gedik, N},
  journal={Physical Review Letters},
  volume={109},
  number={12},
  pages={127401},
  year={2012},
  publisher={APS}
}

@article{lin2008,
  title={Electron-phonon coupling and electron heat capacity of metals under conditions of strong electron-phonon nonequilibrium},
  author={Lin, Zhibin and Zhigilei, Leonid V and Celli, Vittorio},
  journal={Physical Review B},
  volume={77},
  number={7},
  pages={075133},
  year={2008},
  publisher={APS}
}

@article{brouet2008,
  title={Angle-resolved photoemission study of the evolution of band structure and charge density wave properties in {RTe$_3$} ({R= Y, La, Ce, Sm, Gd, Tb, and Dy})},
  author={Brouet, Veronique and Yang, WL and Zhou, XJ and Hussain, Zahid and Moore, RG and He, R and Lu, DH and Shen, ZX and Laverock, J and Dugdale, SB and others},
  journal={Physical Review B},
  volume={77},
  number={23},
  pages={235104},
  year={2008},
  publisher={APS}
}

@article{ru2006,
  title={Thermodynamic and transport properties of {YTe$_3$}, {LaTe$_3$}, and {CeTe$_3$}},
  author={Ru, N and Fisher, IR},
  journal={Physical Review B},
  volume={73},
  number={3},
  pages={033101},
  year={2006},
  publisher={APS}
}

@article{nicholson2019PRB,
  title={Excited-state band mapping and momentum-resolved ultrafast population dynamics in {In/Si} (111) nanowires investigated with {XUV}-based time-and angle-resolved photoemission spectroscopy},
  author={Nicholson, Christopher W and Puppin, Michele and L{\"u}cke, A and Gerstmann, U and Krenz, Marcel and Schmidt, WG and Rettig, Laurenz and Ernstorfer, Ralph and Wolf, Martin},
  journal={Physical Review B},
  volume={99},
  number={15},
  pages={155107},
  year={2019},
  publisher={APS}
}

@article{perfetti2007,
  title={Ultrafast electron relaxation in superconducting {Bi$_2$Sr$_2$CaCu$_2$O$_{8+\delta}$} by time-resolved photoelectron spectroscopy},
  author={Perfetti, L and Loukakos, PA and Lisowski, M and Bovensiepen, U and Eisaki, H and Wolf, M},
  journal={Physical review letters},
  volume={99},
  number={19},
  pages={197001},
  year={2007},
  publisher={APS}
}

@article{johnson2017,
  title={Watching ultrafast responses of structure and magnetism in condensed matter with momentum-resolved probes},
  author={Johnson, Steven L and Savoini, Matteo and Beaud, Paul and Ingold, Gerhard and Staub, Urs and Carbone, Fabrizio and Castiglioni, Luca and Hengsberger, Matthias and Osterwalder, J{\"u}rg},
  journal={Structural Dynamics},
  volume={4},
  number={6},
  pages={061506},
  year={2017},
  publisher={American Crystallographic Association}
}

@article{yamaji1983,
  title={First-order phase transition boundary between superconducting and {SDW} phases in the bechgaard salts},
  author={Yamaji, Kunihiko},
  journal={Journal of the Physical Society of Japan},
  volume={52},
  number={4},
  pages={1361--1372},
  year={1983},
  publisher={The Physical Society of Japan}
}

@article{neugebauer2019,
  title={Optical control of vibrational coherence triggered by an ultrafast phase transition},
  author={Neugebauer, Martin J and Huber, Tim and Savoini, Matteo and Abreu, Elsa and Esposito, Vincent and Kubli, Martin and Rettig, Laurenz and Bothschafter, E and Gr{\"u}bel, Sebastian and Kubacka, Teresa and others},
  journal={Physical Review B},
  volume={99},
  number={22},
  pages={220302},
  year={2019},
  publisher={APS}
}

@article{tao2013,
  title={Anisotropic electron-phonon coupling investigated by ultrafast electron crystallography: Three-temperature model},
  author={Tao, Zhensheng and Han, Tzong-Ru T and Ruan, Chong-Yu},
  journal={Physical Review B},
  volume={87},
  number={23},
  pages={235124},
  year={2013},
  publisher={APS}
}

@article{demsar1999,
  title={Single particle and collective excitations in the one-dimensional charge density wave solid {K}\textsubscript{0.3}{MoO}$_3$ probed in real time by femtosecond spectroscopy},
  author={Demsar, Jure and Biljakovic, Katica and Mihailovic, Dragan},
  journal={Physical review letters},
  volume={83},
  number={4},
  pages={800},
  year={1999},
  publisher={APS}
}

@article{schmitt2008,
  title={Transient electronic structure and melting of a charge density wave in {TbTe$_3$}},
  author={Schmitt, Felix and Kirchmann, Patrick S and Bovensiepen, Uwe and Moore, Rob G and Rettig, Laurenz and Krenz, Marcel and Chu, J-H and Ru, Nancy and Perfetti, Luca and Lu, DH and others},
  journal={Science},
  volume={321},
  number={5896},
  pages={1649--1652},
  year={2008},
  publisher={American Association for the Advancement of Science}
}

@article{huber2014,
  title={Coherent structural dynamics of a prototypical charge-density-wave-to-metal transition},
  author={Huber, Tim and Mariager, Simon O and Ferrer, Andres and Sch{\"a}fer, Hanjo and Johnson, Jeremy A and Gr{\"u}bel, Sebastian and L{\"u}bcke, Andre and Huber, Lucas and Kubacka, Teresa and Dornes, Christian and others},
  journal={Physical review letters},
  volume={113},
  number={2},
  pages={026401},
  year={2014},
  publisher={APS}
}

@article{yusupov2008,
  title={Single-Particle and Collective Mode Couplings Associated with 1-and 2-Directional Electronic Ordering in Metallic {RTe$_3$} ({R= Ho, Dy, Tb})},
  author={Yusupov, RV and Mertelj, T and Chu, J-H and Fisher, IR and Mihailovic, D},
  journal={Physical review letters},
  volume={101},
  number={24},
  pages={246402},
  year={2008},
  publisher={APS}
}

@article{schmitt2011,
  title={Ultrafast electron dynamics in the charge density wave material {TbTe$_3$}},
  author={Schmitt, F and Kirchmann, Patrick S and Bovensiepen, Uwe and Moore, RG and Chu, JH and Lu, DH and Rettig, L and Wolf, M and Fisher, IR and Shen, ZX},
  journal={New Journal of Physics},
  volume={13},
  number={6},
  pages={063022},
  year={2011},
  publisher={IOP Publishing}
}

@article{moore2016,
  title={Ultrafast resonant soft x-ray diffraction dynamics of the charge density wave in {TbTe$_3$}},
  author={Moore, RG and Lee, WS and Kirchman, PS and Chuang, YD and Kemper, AF and Trigo, M and Patthey, L and Lu, DH and Krupin, O and Yi, M and others},
  journal={Physical Review B},
  volume={93},
  number={2},
  pages={024304},
  year={2016},
  publisher={APS}
}

@article{pouget2016,
  title={The Peierls instability and charge density wave in one-dimensional electronic conductors},
  author={Pouget, Jean-Paul},
  journal={Comptes Rendus Physique},
  volume={17},
  number={3-4},
  pages={332--356},
  year={2016},
  publisher={Elsevier}
}

@book{gruner1994,
  title={Density waves in solids},
  author={Gruner, George},
  year={1994},
  publisher={CRC press}
}

@book{motizuki1986,
  title={Structural phase transitions in layered transition metal compounds},
  author={Motizuki, Kazuko},
  year={1986},
  publisher={Springer Science \& Business Media}
}

@article{wandel2020,
  title={Light-enhanced Charge Density Wave Coherence in a High-Temperature Superconductor},
  author={Wandel, S and Boschini, F and Neto, EH and Shen, L and Na, MX and Zohar, S and Wang, Y and Welch, GB and Seaberg, MH and Koralek, JD and others},
  journal={arXiv preprint arXiv:2003.04224},
  year={2020}
}

@article{kogar2020,
  title={Light-induced charge density wave in {LaTe$_3$}},
  author={Kogar, Anshul and Zong, Alfred and Dolgirev, Pavel E and Shen, Xiaozhe and Straquadine, Joshua and Bie, Ya-Qing and Wang, Xirui and Rohwer, Timm and Tung, I-Cheng and Yang, Yafang and others},
  journal={Nature Physics},
  volume={16},
  number={2},
  pages={159--163},
  year={2020},
  publisher={Nature Publishing Group}
}

@article{fausti2011,
  title={Light-induced superconductivity in a stripe-ordered cuprate},
  author={Fausti, Daniele and Tobey, RI and Dean, Nicky and Kaiser, Stefan and Dienst, A and Hoffmann, Matthias C and Pyon, S and Takayama, T and Takagi, H and Cavalleri, Andrea},
  journal={science},
  volume={331},
  number={6014},
  pages={189--191},
  year={2011},
  publisher={American Association for the Advancement of Science}
}

@article{rettig2016,
  title={Persistent order due to transiently enhanced nesting in an electronically excited charge density wave},
  author={Rettig, L and Cort{\'e}s, R and Chu, J-H and Fisher, IR and Schmitt, F and Moore, RG and Shen, Z-X and Kirchmann, PS and Wolf, Martin and Bovensiepen, Uwe},
  journal={Nature communications},
  volume={7},
  number={1},
  pages={1--6},
  year={2016},
  publisher={Nature Publishing Group}
}

@article{perfetti2006,
  title={Time Evolution of the Electronic Structure of {1T-TaS$_2$} through the Insulator-Metal Transition},
  author={Perfetti, Luca and Loukakos, PA and Lisowski, M and Bovensiepen, U and Berger, H and Biermann, S and Cornaglia, PS and Georges, A and Wolf, M},
  journal={Physical review letters},
  volume={97},
  number={6},
  pages={067402},
  year={2006},
  publisher={APS}
}

@article{eichberger2010,
  title={Snapshots of cooperative atomic motions in the optical suppression of charge density waves},
  author={Eichberger, Maximilian and Sch{\"a}fer, Hanjo and Krumova, Marina and Beyer, Markus and Demsar, Jure and Berger, Helmuth and Moriena, Gustavo and Sciaini, Germ{\'a}n and Miller, RJ Dwayne},
  journal={Nature},
  volume={468},
  number={7325},
  pages={799--802},
  year={2010},
  publisher={Nature Publishing Group}
}

@article{stojchevska2014,
  title={Ultrafast switching to a stable hidden quantum state in an electronic crystal},
  author={Stojchevska, L and Vaskivskyi, I and Mertelj, T and Kusar, P and Svetin, D and Brazovskii, S and Mihailovic, D},
  journal={Science},
  volume={344},
  number={6180},
  pages={177--180},
  year={2014},
  publisher={American Association for the Advancement of Science}
}

@article{porer2014,
  title={Non-thermal separation of electronic and structural orders in a persisting charge density wave},
  author={Porer, Michael and Leierseder, Ursula and M{\'e}nard, J-M and Dachraoui, Hatem and Mouchliadis, L and Perakis, IE and Heinzmann, Ulrich and Demsar, J and Rossnagel, K and Huber, Rupert},
  journal={Nature materials},
  volume={13},
  number={9},
  pages={857--861},
  year={2014},
  publisher={Nature Publishing Group}
}

@article{zhang2016,
  title={Cooperative photoinduced metastable phase control in strained manganite films},
  author={Zhang, Jingdi and Tan, Xuelian and Liu, Mengkun and Teitelbaum, Samuel W and Post, Kirk W and Jin, Feng and Nelson, Keith A and Basov, DN and Wu, Wenbin and Averitt, Richard D},
  journal={Nature materials},
  volume={15},
  number={9},
  pages={956--960},
  year={2016},
  publisher={Nature Publishing Group}
}

@article{maschek2015,
  title={Wave-vector-dependent electron-phonon coupling and the charge-density-wave transition in TbT e 3},
  author={Maschek, M and Rosenkranz, S and Heid, R and Said, AH and Giraldo-Gallo, P and Fisher, IR and Weber, F},
  journal={Physical Review B},
  volume={91},
  number={23},
  pages={235146},
  year={2015},
  publisher={APS}
}

@article{tsuji2013,
  title={Nonthermal antiferromagnetic order and nonequilibrium criticality in the Hubbard model},
  author={Tsuji, Naoto and Eckstein, Martin and Werner, Philipp},
  journal={Physical review letters},
  volume={110},
  number={13},
  pages={136404},
  year={2013},
  publisher={APS}
}

@article{gerasimenko2019,
  title={Intertwined chiral charge orders and topological stabilization of the light-induced state of a prototypical transition metal dichalcogenide},
  author={Gerasimenko, Yaroslav A and Karpov, Petr and Vaskivskyi, Igor and Brazovskii, Serguei and Mihailovic, Dragan},
  journal={npj Quantum Materials},
  volume={4},
  number={1},
  pages={1--9},
  year={2019},
  publisher={Nature Publishing Group}
}

@article{beaud2014,
  title={A time-dependent order parameter for ultrafast photoinduced phase transitions},
  author={Beaud, P and Caviezel, A and Mariager, SO and Rettig, L and Ingold, G and Dornes, C and Huang, SW and Johnson, JA and Radovic, M and Huber, T and others},
  journal={Nature materials},
  volume={13},
  number={10},
  pages={923--927},
  year={2014},
  publisher={Nature Publishing Group}
}

@article{ru2008,
  title={Effect of chemical pressure on the charge density wave transition in rare-earth tritellurides {RTe$_3$}},
  author={Ru, N and Condron, CL and Margulis, GY and Shin, KY and Laverock, J and Dugdale, SB and Toney, MF and Fisher, IR},
  journal={Physical Review B},
  volume={77},
  number={3},
  pages={035114},
  year={2008},
  publisher={APS}
}

@article{trigo2019,
  title={Coherent order parameter dynamics in {SmTe$_3$}},
  author={Trigo, Mariano and Giraldo-Gallo, P and Kozina, ME and Henighan, T and Jiang, MP and Liu, H and Clark, JN and Chollet, M and Glownia, JM and Zhu, D and others},
  journal={Physical Review B},
  volume={99},
  number={10},
  pages={104111},
  year={2019},
  publisher={APS}
}

@article{yusupov2010,
  title={Coherent dynamics of macroscopic electronic order through a symmetry breaking transition},
  author={Yusupov, Roman and Mertelj, Tomaz and Kabanov, Viktor V and Brazovskii, Serguei and Kusar, Primoz and Chu, Jiun-Haw and Fisher, Ian R and Mihailovic, Dragan},
  journal={Nature Physics},
  volume={6},
  number={9},
  pages={681--684},
  year={2010},
  publisher={Nature Publishing Group}
}

@article{schaefer2014,
  title={Collective modes in quasi-one-dimensional charge-density wave systems probed by femtosecond time-resolved optical studies},
  author={Schaefer, Hanjo and Kabanov, Viktor V and Demsar, Jure},
  journal={Physical Review B},
  volume={89},
  number={4},
  pages={045106},
  year={2014},
  publisher={APS}
}

@article{arguello2014,
  title={Visualizing the charge density wave transition in {2H-NbSe$_2$} in real space},
  author={Arguello, CJ and Chockalingam, SP and Rosenthal, EP and Zhao, L and Guti{\'e}rrez, C and Kang, JH and Chung, WC and Fernandes, RM and Jia, S and Millis, AJ and others},
  journal={Physical Review B},
  volume={89},
  number={23},
  pages={235115},
  year={2014},
  publisher={APS}
}

@article{fang2019,
  title={Disorder-induced suppression of charge density wave order: {STM} study of {Pd}-intercalated {ErTe$_3$}},
  author={Fang, Alan and Straquadine, Joshua AW and Fisher, Ian R and Kivelson, Steven A and Kapitulnik, Aharon},
  journal={Physical Review B},
  volume={100},
  number={23},
  pages={235446},
  year={2019},
  publisher={APS}
}

@article{tomeljak2009,
  title={Dynamics of photoinduced charge-density-wave to metal phase transition in {K$_{0.3}$MoO$_3$}},
  author={Tomeljak, Andrej and Schaefer, Hanjo and St{\"a}dter, David and Beyer, Markus and Biljakovic, Katica and Demsar, Jure},
  journal={Physical review letters},
  volume={102},
  number={6},
  pages={066404},
  year={2009},
  publisher={APS}
}

@book{goldenfeld1992,
  title={Lectures on phase transitions and the renormalization group},
  author={Goldenfeld, Nigel},
  year={1992},
  publisher={CRC Press}
}

@article{zong2019evidence,
  title={Evidence for topological defects in a photoinduced phase transition},
  author={Zong, Alfred and Kogar, Anshul and Bie, Ya-Qing and Rohwer, Timm and Lee, Changmin and Baldini, Edoardo and Erge{\c{c}}en, Emre and Yilmaz, Mehmet B and Freelon, Byron and Sie, Edbert J and others},
  journal={Nature Physics},
  volume={15},
  number={1},
  pages={27--31},
  year={2019},
  publisher={Nature Publishing Group}
}

@article{mohr2011,
  title={Nonthermal melting of a charge density wave in {TiSe$_2$}},
  author={M{\"o}hr-Vorobeva, Ekaterina and Johnson, Steven L and Beaud, Paul and Staub, Urs and De Souza, R and Milne, Chris and Ingold, Gerhard and Demsar, Jure and Sch{\"a}fer, Hanjo and Titov, A},
  journal={Physical review letters},
  volume={107},
  number={3},
  pages={036403},
  year={2011},
  publisher={APS}
}

@article{endres2012,
  title={The ‘Higgs’ amplitude mode at the two-dimensional superfluid/Mott insulator transition},
  author={Endres, Manuel and Fukuhara, Takeshi and Pekker, David and Cheneau, Marc and Schau$\beta$, Peter and Gross, Christian and Demler, Eugene and Kuhr, Stefan and Bloch, Immanuel},
  journal={Nature},
  volume={487},
  number={7408},
  pages={454--458},
  year={2012},
  publisher={Nature Publishing Group}
}

@article{matsunaga2013,
  title={Higgs amplitude mode in the {BCS} superconductors {Nb$_{1-x}$Ti$_x$N} induced by terahertz pulse excitation},
  author={Matsunaga, Ryusuke and Hamada, Yuki I and Makise, Kazumasa and Uzawa, Yoshinori and Terai, Hirotaka and Wang, Zhen and Shimano, Ryo},
  journal={Physical review letters},
  volume={111},
  number={5},
  pages={057002},
  year={2013},
  publisher={APS}
}

@article{torchinsky2013,
  title={Fluctuating charge-density waves in a cuprate superconductor},
  author={Torchinsky, Darius H and Mahmood, Fahad and Bollinger, Anthony T and Bo{\v{z}}ovi{\'c}, Ivan and Gedik, Nuh},
  journal={Nature materials},
  volume={12},
  number={5},
  pages={387--391},
  year={2013},
  publisher={Nature Publishing Group}
}

@article{dolgirev2020self,
  title={Self-similar dynamics of order parameter fluctuations in pump-probe experiments},
  author={Dolgirev, Pavel E and Michael, Marios H and Zong, Alfred and Gedik, Nuh and Demler, Eugene},
  journal={Physical Review B},
  volume={101},
  number={17},
  pages={174306},
  year={2020},
  publisher={APS}
}

@article{storeck2019hot,
  title={Structural dynamics of incommensurate charge-density waves tracked by ultrafast low-energy electron diffraction},
  author={Storeck, Gero and Horstmann, Jan Gerrit and Diekmann, Theo and Vogelgesang, Simon and von Witte, Gevin and Yalunin, SV and Rossnagel, Kai and Ropers, Claus},
  journal={Structural Dynamics},
  volume={7},
  number={3},
  pages={034304},
  year={2020},
  publisher={American Crystallographic Association}
}

@article{vaskivskyi2015controlling,
  title={Controlling the metal-to-insulator relaxation of the metastable hidden quantum state in {1T-TaS$_2$}},
  author={Vaskivskyi, Igor and Gospodaric, Jan and Brazovskii, Serguei and Svetin, Damjan and Sutar, Petra and Goreshnik, Evgeny and Mihailovic, Ian A and Mertelj, Tomaz and Mihailovic, Dragan},
  journal={Science advances},
  volume={1},
  number={6},
  pages={e1500168},
  year={2015},
  publisher={American Association for the Advancement of Science}
}

@article{mitrano2016possible,
  title={Possible light-induced superconductivity in {K$_3$C$_{60}$} at high temperature},
  author={Mitrano, Matteo and Cantaluppi, Alice and Nicoletti, Daniele and Kaiser, Stefan and Perucchi, A and Lupi, S and Di Pietro, P and Pontiroli, D and Ricc{\`o}, M and Clark, Stephen R and others},
  journal={Nature},
  volume={530},
  number={7591},
  pages={461--464},
  year={2016},
  publisher={Nature Publishing Group}
}

@article{cavalleri2018photo,
  title={Photo-induced superconductivity},
  author={Cavalleri, Andrea},
  journal={Contemporary Physics},
  volume={59},
  number={1},
  pages={31--46},
  year={2018},
  publisher={Taylor \& Francis}
}

@article{monney2016revealing,
  title={Revealing the role of electrons and phonons in the ultrafast recovery of charge density wave correlations in {1T- TiSe$_2$}},
  author={Monney, Claude and Puppin, Michele and Nicholson, CW and Hoesch, M and Chapman, RT and Springate, E and Berger, H and Magrez, A and Cacho, C and Ernstorfer, Ralph and others},
  journal={Physical Review B},
  volume={94},
  number={16},
  pages={165165},
  year={2016},
  publisher={APS}
}

@article{singer2016photoinduced,
  title={Photoinduced enhancement of the charge density wave amplitude},
  author={Singer, A and Patel, SKK and Kukreja, R and Uhl{\'{i}}{\v{r}}, V and Wingert, J and Festersen, S and Zhu, D and Glownia, JM and Lemke, HT and Nelson, S and others},
  journal={Physical review letters},
  volume={117},
  number={5},
  pages={056401},
  year={2016},
  publisher={APS}
}

@article{nicholson2018beyond,
  title={Beyond the molecular movie: Dynamics of bands and bonds during a photoinduced phase transition},
  author={Nicholson, Christopher W and L{\"u}cke, Andreas and Schmidt, Wolf Gero and Puppin, Michele and Rettig, Laurenz and Ernstorfer, Ralph and Wolf, Martin},
  journal={Science},
  volume={362},
  number={6416},
  pages={821--825},
  year={2018},
  publisher={American Association for the Advancement of Science}
}

@article{wall2018ultrafast,
  title={Ultrafast disordering of vanadium dimers in photoexcited VO2},
  author={Wall, Simon and Yang, Shan and Vidas, Luciana and Chollet, Matthieu and Glownia, James M and Kozina, Michael and Katayama, Tetsuo and Henighan, Thomas and Jiang, Mason and Miller, Timothy A and others},
  journal={Science},
  volume={362},
  number={6414},
  pages={572--576},
  year={2018},
  publisher={American Association for the Advancement of Science}
}

@article{tengdin2018critical,
  title={Critical behavior within 20 fs drives the out-of-equilibrium laser-induced magnetic phase transition in nickel},
  author={Tengdin, Phoebe and You, Wenjing and Chen, Cong and Shi, Xun and Zusin, Dmitriy and Zhang, Yingchao and Gentry, Christian and Blonsky, Adam and Keller, Mark and Oppeneer, Peter M and others},
  journal={Science advances},
  volume={4},
  number={3},
  pages ={9744},
  year={2018},
  publisher={American Association for the Advancement of Science}
}

@article{nicholson2016ultrafast,
  title={Ultrafast spin density wave transition in chromium governed by thermalized electron gas},
  author={Nicholson, CW and Monney, C and Carley, R and Frietsch, Bj{\"o}rn and Bowlan, J and Weinelt, Martin and Wolf, Martin},
  journal={Physical review letters},
  volume={117},
  number={13},
  pages={136801},
  year={2016},
  publisher={APS}
}

@article{yang2020bypassing,
  title={Bypassing the structural bottleneck in the ultrafast melting of electronic order},
  author={Yang, LX and Rohde, G and Hanff, K and Stange, A and Xiong, R and Shi, J and Bauer, M and Rossnagel, K},
  journal={Physical Review Letters},
  volume={125},
  number={26},
  pages={266402},
  year={2020},
  publisher={APS}
}

@article{overhauser1971observability,
  title={Observability of charge-density waves by neutron diffraction},
  author={Overhauser, AW},
  journal={Physical Review B},
  volume={3},
  number={10},
  pages={3173},
  year={1971},
  publisher={APS}
}

@article{vogelgesang2018phase,
  title={Phase ordering of charge density waves traced by ultrafast low-energy electron diffraction},
  author={Vogelgesang, S and Storeck, G and Horstmann, JG and Diekmann, T and Sivis, M and Schramm, S and Rossnagel, K and Sch{\"a}fer, S and Ropers, C},
  journal={Nature Physics},
  volume={14},
  number={2},
  pages={184--190},
  year={2018},
  publisher={Nature Publishing Group}
}

@article{waldecker2016electron,
  title={Electron-phonon coupling and energy flow in a simple metal beyond the two-temperature approximation},
  author={Waldecker, Lutz and Bertoni, Roman and Ernstorfer, Ralph and Vorberger, Jan},
  journal={Physical Review X},
  volume={6},
  number={2},
  pages={021003},
  year={2016},
  publisher={APS}
}

@article{trigo2020formation,
  title={Ultrafast formation of domain walls of a charge density wave in {SmTe$_3$}},
  author={Trigo, M and Giraldo-Gallo, P and Clark, JN and Kozina, ME and Henighan, T and Jiang, MP and Chollet, M and Fisher, IR and Glownia, JM and Katayama, T and others},
  journal={Physical Review B},
  volume={103},
  number={5},
  pages={054109},
  year={2021},
  publisher={APS}
}

@article{hellmann2012time,
  title={Time-domain classification of charge-density-wave insulators},
  author={Hellmann, S and Rohwer, T and Kall{\"a}ne, M and Hanff, K and Sohrt, C and Stange, A and Carr, A and Murnane, MM and Kapteyn, HC and Kipp, L and others},
  journal={Nature communications},
  volume={3},
  number={1},
  pages={1--8},
  year={2012},
  publisher={Nature Publishing Group}
}

@article{sohrt2014fast,
  title={How fast can a Peierls--Mott insulator be melted?},
  author={Sohrt, C and Stange, A and Bauer, M and Rossnagel, K},
  journal={Faraday discussions},
  volume={171},
  pages={243--257},
  year={2014},
  publisher={Royal Society of Chemistry}
}

@article{rettig2016itinerant,
  title={Itinerant and localized magnetization dynamics in antiferromagnetic {Ho}},
  author={Rettig, Laurenz and Dornes, Christian and Thielemann-K{\"u}hn, Nele and Pontius, Niko and Zabel, Hartmut and Schlagel, Deborah L and Lograsso, Tommaso A and Chollet, Matthieu and Robert, Aymeric and Sikorski, Marcin and others},
  journal={Physical review letters},
  volume={116},
  number={25},
  pages={257202},
  year={2016},
  publisher={APS}
}

@article{laverock2005fermi,
  title={Fermi surface nesting and charge-density wave formation in rare-earth tritellurides},
  author={Laverock, J and Dugdale, SB and Major, Zs and Alam, MA and Ru, N and Fisher, IR and Santi, G and Bruno, E},
  journal={Physical Review B},
  volume={71},
  number={8},
  pages={085114},
  year={2005},
  publisher={APS}
}

@article{maschek2018competing,
  title={Competing soft phonon modes at the charge-density-wave transitions in DyT e 3},
  author={Maschek, M and Zocco, DA and Rosenkranz, S and Heid, R and Said, AH and Alatas, A and Walmsley, P and Fisher, IR and Weber, F},
  journal={Physical Review B},
  volume={98},
  number={9},
  pages={094304},
  year={2018},
  publisher={APS}
}

@dataset{data2020,
  author       = {Maklar, J. and
                  Windsor, Y. W. and
                  Nicholson, C.W. and
                  Puppin, M. and
                  Walmsley, P. and
                  Esposito, V. and
                  Porer, M. and
                  Rittmann, J. and
                  Leuenberger, D. and
                  Kubli, M. and
                  Savoini, M. and
                  Abreu, E. and
                  Johnson, S.L. and
                  Beaud, P. and
                  Ingold, G. and
                  Staub, U. and
                  Fisher, I.R. and
                  Ernstorfer, R. and
                  Wolf, M. and
                  Rettig, L.},
  title        = {{Time- and angle-resolved photoemission 
                   spectroscopy data and time-resolved X-ray
                   diffraction data of TbTe3}},
  month        = oct,
  year         = 2020,
  publisher    = {Zenodo},
  version      = {1.0.0},
  doi          = {10.5281/zenodo.4106272},
  url          = {https://doi.org/10.5281/zenodo.4106272}
}

\end{document}